\documentclass[english]{IEEEtran}
\usepackage[T1]{fontenc}
\usepackage[latin9]{inputenc}
\usepackage{color}
\usepackage{float}
\usepackage{mathrsfs}
\usepackage{amsmath}
\usepackage{amsthm}
\usepackage{amssymb}
\usepackage{graphicx}
\usepackage{setspace}

\makeatletter

\providecommand{\tabularnewline}{\\}
\floatstyle{ruled}
\newfloat{algorithm}{tbp}{loa}
\providecommand{\algorithmname}{Algorithm}
\floatname{algorithm}{\protect\algorithmname}

\theoremstyle{plain}
\newtheorem{thm}{\protect\theoremname}
\theoremstyle{definition}
\newtheorem{defn}[thm]{\protect\definitionname}
\theoremstyle{plain}
\newtheorem{lem}[thm]{\protect\lemmaname}

\usepackage{cite}
\usepackage{times}
\usepackage{URL}

\usepackage{algorithmic}

\usepackage{subfig}

\makeatother

\usepackage{babel}
\providecommand{\definitionname}{Definition}
\providecommand{\lemmaname}{Lemma}
\providecommand{\theoremname}{Theorem}

\begin{document}
\title{Subspace Constrained Variational Bayesian Inference for Structured
Compressive Sensing with a Dynamic Grid}
\author{{\normalsize{}An Liu, }\textit{\normalsize{}Senior Member, IEEE}{\normalsize{},
Yufan Zhou, and Wenkang Xu}\thanks{An Liu, Yufan Zhou, and Wenkang Xu are with the College of Information
Science and Electronic Engineering, Zhejiang University, Hangzhou
310027, China (email: anliu@zju.edu.cn).}}
\maketitle
\begin{abstract}
We investigate the problem of recovering a structured sparse signal
from a linear observation model with an uncertain dynamic grid in
the sensing matrix. The state-of-the-art expectation maximization
based compressed sensing (EM-CS) methods, such as turbo compressed
sensing (Turbo-CS) and turbo variational Bayesian inference (Turbo-VBI),
have a relatively slow convergence speed due to the double-loop iterations
between the E-step and M-step. Moreover, each inner iteration in the
E-step involves a high-dimensional matrix inverse in general, which
is unacceptable for problems with large signal dimensions or real-time
calculation requirements. Although there are some attempts to avoid
the high-dimensional matrix inverse by majorization minimization,
the convergence speed and accuracy are often sacrificed. To better
address this problem, we propose an alternating estimation framework
based on a novel subspace constrained VBI (SC-VBI) method, in which
the high-dimensional matrix inverse is replaced by a low-dimensional
subspace constrained matrix inverse (with the dimension equal to the
sparsity level). We further prove the convergence of the SC-VBI to
a stationary solution of the Kullback-Leibler divergence minimization
problem. Simulations demonstrate that the proposed SC-VBI algorithm
can achieve a much better tradeoff between complexity per iteration,
convergence speed, and performance compared to the state-of-the-art
algorithms.
\end{abstract}

\begin{IEEEkeywords}
Subspace constrained variational Bayesian inference (SC-VBI), structured
sparse signal recovery, dynamic grid.

\thispagestyle{empty}
\end{IEEEkeywords}

\section{Introduction}

The problem of structured compressive sensing (CS) with dynamic grids
has attracted a lot of research due to its wide applications in signal
processing and wireless communications \cite{OGSBI,Zhou_Offgrid,Dai_MIMOCE}.
The goal of this problem is to recover a structured sparse signal
$\boldsymbol{x}=[x_{1},...,x_{N}]^{T}\in\mathbb{C}^{N\times1}$ from
significantly fewer measurements $\boldsymbol{y}\in\mathbb{C}^{M\times1}$
(i.e., $M\ll N$) under the following linear observation model:
\begin{equation}
\boldsymbol{\boldsymbol{y}}=\mathbf{A}\left(\boldsymbol{\theta}\right)\boldsymbol{x}+\boldsymbol{w},\label{eq:linear model}
\end{equation}
where $\mathbf{A}\left(\boldsymbol{\theta}\right)\in\mathbb{C}^{M\times N}$
is the sensing matrix with dynamic grid parameters $\boldsymbol{\theta}$,
${\color{red}{\color{black}\boldsymbol{w}\in\mathbb{C}^{M\times1}}}$
is the noise vector with independent Gaussian entries.

In the standard CS problem, $\boldsymbol{\theta}$ is chosen to be
a pre-determined fixed grid and $\boldsymbol{x}$ is an i.i.d. sparse
signal. However, in practice, the true grid parameters $\boldsymbol{\theta}$
usually do not lie exactly on the pre-determined fixed grid points.
For example, in massive MIMO channel estimation, the grid parameters
$\boldsymbol{\theta}$ correspond to the angles of the active channel
paths. If we use a fixed angle grid, the estimation accuracy of the
angles $\boldsymbol{\theta}$ will be limited by the grid resolution,
which will also cause energy leakage and degrade the channel estimation
performance \cite{OGSBI}. Future 6G wireless systems are expected
to also provide sensing services, and it is important to achieve ``super
resolution'' estimation of the angle/delay parameters to realize
high-accuracy sensing and localization \cite{LiuFan_ISAC_survey,DeLima_Sensing_survey}.
In this case, we also need to adjust the grid to achieve more accurate
estimation of the grid parameters $\boldsymbol{\theta}$ and alleviate
the energy leakage effect. Moreover, in many practical applications,
the sparse signal $\boldsymbol{x}$ usually has structured sparsity
that cannot be modeled easily by an i.i.d. prior. Therefore, it is
very important to consider the structured CS problem with a dynamic
grid.

The state-of-the-art methods to solve the structured CS problem with
a dynamic grid are those expectation maximization based compressed
sensing (EM-CS) methods, which iterates between the E-step and M-step.
In the E-step, the turbo approach is often applied to compute a Bayesian
estimation of $\boldsymbol{x}$ with a structured sparse prior by
exchanging extrinsic messages between two modules until convergence.
Specifically, one module (say Module A) performs Bayesian inference
over the observation model and another module (say Module B) performs
message passing over the structured sparse prior. Various Turbo-based
algorithms have been proposed for the E-step, such as turbo approximate
message passing (Turbo-AMP) \cite{Som_TurboAMP}, turbo compressive
sensing (Turbo-CS) \cite{Yuan_TurboCS,LiuAn_TurboOAMP}, and turbo
variational Bayesian inference (Turbo-VBI) \cite{LiuAn_CE_Turbo_VBI,LiuAn_directloc_vehicles},
whose Module A is the AMP estimator, the linear minimum mean square
error (LMMSE) estimator, and the VBI estimator, respectively. In the
M-step, a surrogate function of the likelihood function for the grid
parameters is firstly constructed from the Bayesian estimation of
$\boldsymbol{x}$ obtained in the E-step and then the gradient ascent
method is used to update the dynamic grid parameters by maximizing
the surrogate function.

Although the EM-CS methods are shown to achieve a good performance
in various applications \cite{Dai_MIMOCE,LiuAn_TurboOAMP,Lian_dynamic_sparsity,Huangzhe_TurboSBI},
there are still several drawbacks. Firstly, the EM-CS methods contain
two-loop of iterations, namely, the inner iterations in the E-step
between Module A and Module B and the outer iteration between the
E-step and M-step. Secondly, the Module A in the E-step involves an
$N$-dimensional matrix inverse in general, which has a high computational
complexity for large-scale problems (i.e., when $N$ is large). There
are a few exceptions, e.g., when the sensing matrix $\mathbf{A}\left(\boldsymbol{\theta}\right)$
has i.i.d. entries or is partially orthogonal (i.e., the rows of the
sensing matrix are mutually orthogonal), the matrix inverse can be
avoided in the E-step of Turbo-AMP and Turbo-CS, respectively. However,
this is not the case for many practical applications, especially when
the sensing matrix contains dynamic grid parameters $\boldsymbol{\theta}$.
It is impossible for the sensing matrix $\mathbf{A}\left(\boldsymbol{\theta}\right)$
to satisfy the above good properties for different values of $\boldsymbol{\theta}$.
Finally, the M-step is based on maximizing a surrogate likelihood
function of $\boldsymbol{\theta}$, making it difficult to choose
a good step size using e.g., backtrack line search, since a good step
size for the surrogate likelihood function is not necessarily good
for the true likelihood function (as observed in many experiments).
As such, the EM-CS methods often need a relatively large number of
inner iterations to converge to a good accuracy and the complexity
per inner iteration is also high especially for large-scale problems.

Recently, there are a few works attempting to avoid the high-dimensional
matrix inverse. For example, in \cite{Duan_IFSBL,Xu_Turbo-IFVBI},
the authors proposed an inverse-free VBI (IF-VBI) algorithm that avoids
the matrix inverse by maximizing a relaxed evidence lower bound (ELBO).
In a very recent work \cite{Xu_SLA_VBI}, an inverse-free successive
linear approximation VBI (IFSLA-VBI) method was developed, where the
matrix inverse is avoided by solving a quadratic minimization problem
based on the majorization minimization (MM) framework. Although the
MM methods in \cite{Duan_IFSBL,Xu_Turbo-IFVBI,Xu_SLA_VBI} completely
avoided the matrix inverse, they added another loop of MM iterations
to approximate the high-dimensional matrix inverse and thus the overall
convergence speed is expected to be slow. Moreover, the IF-VBI method
needs a singular value decomposition (SVD) operation additionally
in order to construct the ELBO, and the IFSLA-VBI method adds some
additional calculations due to the successive linear expansion of
the observation model and consideration of Bayesian inference for
the non-linear dynamic parameters $\boldsymbol{\theta}$.

Another line of works have proposed an atomic norm minimization approach
to achieve gridless CS. In \cite{Tang_ANM,Yang_ANM}, the authors
showed that the sparse signal and sensing parameters can be exactly
recovered with high probability via atomic norm minimization, provided
the sensing parameters are well separated. However, atomic norm minimization
involves semi-definite programming, which has a much higher complexity
than the EM-CS method. Moreover, it is difficult to incorporate the
more complicated structured sparsity in the framework of atomic norm
minimization.

In this paper, we propose an alternating estimation framework based
on a novel subspace constrained VBI (SC-VBI) method to overcome the
drawbacks of the existing methods. The main contributions are summarized
below.
\begin{itemize}
\item \textbf{An alternating estimation framework: }We propose an alternating
estimation framework that iterates among the SC-VBI module, the grid
estimation (GE) module, and the structured sparse inference (SSI)
module. For fixed grid parameter $\boldsymbol{\theta}$ and extrinsic
message from the SSI module, the SC-VBI module computes a Bayesian
estimation of $\boldsymbol{x}$ based on the observation model. For
fixed MMSE estimator of $\boldsymbol{x}$ output from the SC-VBI module,
the GE module refines the dynamic gird by directly maximizing the
likelihood function based on the gradient ascent method with backtracking
line search. Finally, for fixed extrinsic message from the SC-VBI
module, the SSI module performs sum-product message passing over the
structured sparse prior to exploit the specific sparse structures
in practical applications.
\item \textbf{Subspace constrained VBI (SC-VBI):} The key motivation for
SC-VBI comes from the following observation. Suppose the sparsity
level of $\boldsymbol{x}$ is $S\ll N$. If we know the support of
the sparse signal (i.e., the index set $\mathcal{S}$ to indicate
the locations of the non-zero elements in the sparse signal $\boldsymbol{x}$),
we can simply omit the zero elements in $\boldsymbol{x}$ and perform
LMMSE to recover $\boldsymbol{x}$ with only a $S$-dimensional matrix
inverse. Therefore, let $\mathbf{W}_{x}$ denote the $N$-dimensional
matrix that needs to be inverted in the VBI. We can first estimate
the support $\hat{\mathcal{S}}$ and then set all off-diagonal elements
that are not in the rows/columns indexed by $\hat{\mathcal{S}}$ to
zero. Then the resulting matrix inverse $\mathbf{W}_{x}^{-1}$ can
be completed by only calculating the inverse of a $S$-dimensional
sub-matrix of $\mathbf{W}_{x}$ restricted to the subspace indexed
by $\hat{\mathcal{S}}$. To fix the error caused by using the estimated
support $\hat{\mathcal{S}}$ instead of the true support $\mathcal{S}$,
we further apply $B_{x}\geq1$ gradient update to refine the posterior
mean of $\boldsymbol{x}$. Finally, a robust design is also introduced
to ensure the fast convergence of the SC-VBI.
\item \textbf{Theoretical Convergence Analysis for SC-VBI:} Despite the
error caused by using the estimated support $\hat{\mathcal{S}}$ and
a finite number of gradient updates, we show that the proposed SC-VBI
converges to a stationary solution of the original VBI problem (i.e.,
the KL-divergence (KLD) minimization problem). 
\end{itemize}
The rest of the paper is organized as follows. In Section \ref{sec:System-Model},
we introduce a three-layer sparse prior model and formulate the structured
CS problem with a dynamic grid. The proposed alternating estimation
framework is presented in Section \ref{sec:The-proposed-Alternating}.
The SC-VBI algorithm and its convergence analysis are presented in
Section \ref{sec:Turbo-VBI-Algorithm}. The proposed algorithm are
applied to solve a massive MIMO channel estimation problem in Section
\ref{sec:Applications}. Finally, the conclusion is given in Section
\ref{sec:Conlusion}.

\textit{Notation:} Lowercase boldface letters denote vectors and uppercase
boldface letters denote matrices. $\left(\cdot\right)^{-1}$, $\left(\cdot\right)^{T}$,
$\left(\cdot\right)^{H}$, $\left\Vert \cdot\right\Vert $, and $\left\langle \cdot\right\rangle $
are used to represent the inverse, transpose, conjugate transpose,
$\ell_{2}\textrm{-norm}$, and expectation operations, respectively.
$\mathbf{I}_{M}$ denotes the $M\times M$ dimensional identity matrix.
Let $\mathfrak{Re}\left\{ \cdot\right\} $ denote the real part of
the complex argument. For a vector $\boldsymbol{x}\in\mathbb{C}^{N}$
and a given index set $\mathcal{S}\subseteq\left\{ 1,...,N\right\} $,
$\left|\mathcal{S}\right|$ denotes its cardinality, $\boldsymbol{x}_{\mathcal{S}}\in\mathbb{C}^{\left|\mathcal{S}\right|\times1}$
denotes the subvector consisting of the elements of $\boldsymbol{x}$
indexed by the set $\mathcal{S}$. $\text{diag}\left(\boldsymbol{x}\right)$
denotes a block diagonal matrix with $\boldsymbol{x}$ as the diagonal
elements. $\mathcal{CN}\left(\boldsymbol{x};\boldsymbol{\mu},\mathbf{\Sigma}\right)$
represents a complex Gaussian distribution with mean $\boldsymbol{\mu}$
and covariance matrix $\mathbf{\Sigma}$. $\Gamma\left(x;a,b\right)$
represents a Gamma distribution with shape parameter $a$ and rate
parameter $b$. Finally, $x=\Theta(a)$ for $a>0$ denotes that $\exists k_{1},k_{2}>0$,
such that $k_{2}\cdot a\leq x\leq k_{1}\cdot a$.

\section{Structured Sparsity Model and Problem Formulation\label{sec:System-Model}}

\subsection{Three-layer Sparse Prior Model}

The probability model for structured sparsity provides the foundation
for exploiting the specific sparse structures. A good sparse probability
model should satisfy the following criteria: it is flexible to capture
the sparse structure of the sparse signal; it is robust w.r.t. the
imperfect prior information in practice; and it is tractable to enable
low-complexity and high-performance algorithm design. In the following,
we shall briefly introduce the three-layer sparse prior model proposed
in \cite{LiuAn_CE_Turbo_VBI}, which has been verified to meet all
these criteria in \cite{LiuAn_CE_Turbo_VBI,LiuAn_directloc_vehicles}.

\begin{figure}
\centering{}\includegraphics[width=75mm]{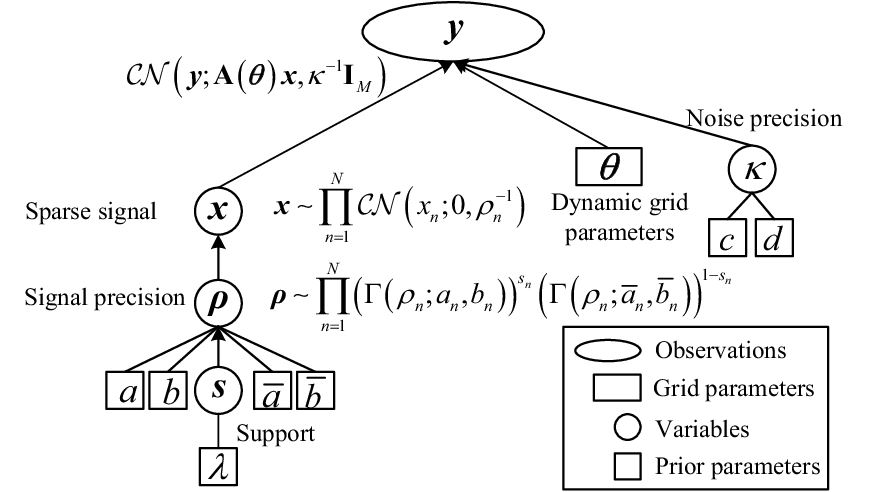}\caption{\label{fig:3LHS-Model}Three-layer hierarchical structured sparse
prior.}
\end{figure}

In the three-layer sparse prior, a support vector $\boldsymbol{s}\triangleq\left[s_{1},\ldots,s_{N}\right]^{T}\in\left\{ 0,1\right\} ^{N}$
is introduced to indicate whether the $n$-th element $x_{n}$ in
$\boldsymbol{x}$ is active ($s_{n}=1$) or inactive ($s_{n}=0$).
Specifically, let $\boldsymbol{\rho}=\left[\rho_{1},...,\rho_{N}\right]^{T}$
denote the precision vector of $\boldsymbol{x}$ (i.e., $1/\rho_{n}$
denotes the variance of $x_{n}$). Then the joint distribution of
$\boldsymbol{x}$, $\boldsymbol{\rho}$, and $\boldsymbol{s}$ can
be expressed as
\begin{equation}
p\left(\boldsymbol{x},\boldsymbol{\rho},\boldsymbol{s}\right)=\underbrace{p\left(\boldsymbol{s}\right)}_{\textrm{Support}}\underbrace{p\left(\boldsymbol{\rho}\mid\boldsymbol{s}\right)}_{\textrm{Precision}}\underbrace{p\left(\boldsymbol{x}\mid\boldsymbol{\rho}\right)}_{\textrm{Sparse\ signal}},\label{eq:p(x,rou,s)}
\end{equation}
as illustrated in Fig. \ref{fig:3LHS-Model}. In the following, we
detail the probability model for each variable.

The prior distribution $p\left(\boldsymbol{s}\right)$ of the support
vector is used to capture the structured sparsity in specific applications.
For example, to capture an independent sparse structure, we can set
\begin{equation}
p\left(\boldsymbol{s}\right)=\prod_{n=1}^{N}\left(\lambda_{n}\right)^{s_{n}}\left(1-\lambda_{n}\right)^{1-s_{n}},
\end{equation}
where $\lambda_{n}$ is the sparsity ratio. In \cite{LiuAn_CE_Turbo_VBI},
$p\left(\boldsymbol{s}\right)$ is chosen to be a Markov chain to
model the clustered scattering environment in massive MIMO channel.

The conditional probability $p\left(\boldsymbol{\rho}\mid\boldsymbol{s}\right)$
is given by
\begin{align}
p\left(\boldsymbol{\rho}\mid\boldsymbol{s}\right) & =\prod_{n=1}^{N}\left(\Gamma\left(\rho_{n};a_{n},b_{n}\right)\right)^{s_{n}}\left(\Gamma\left(\rho_{n};\overline{a}_{n},\overline{b}_{n}\right)\right)^{1-s_{n}},\label{eq:ruoconds}
\end{align}
where $\Gamma\left(\rho;a,b\right)$ is a Gamma hyper-prior with shape
parameter $a$ and rate parameter $b$. When $s_{n}=1$, the variance
$1/\rho_{n}$ of $x_{n}$ is $\Theta\left(1\right)$, and thus the
shape and rate parameters $a_{n},b_{n}$ should be chosen such that
$\frac{a_{n}}{b_{n}}=\mathbb{E}\left[\rho_{n}\right]=\Theta\left(1\right)$.
On the other hand, when $s_{n}=0$, $x_{n}$ is close to zero, and
thus the shape and rate parameters $\overline{a}_{n},\overline{b}_{n}$
should be chosen to satisfy $\frac{\overline{a}_{n}}{\overline{b}_{n}}=\mathbb{E}\left[\rho_{n}\right]\gg1$.

The conditional probability $p\left(\boldsymbol{x}\mid\boldsymbol{\rho}\right)$
for the sparse signal is assumed to have a product form $p\left(\boldsymbol{x}\mid\boldsymbol{\rho}\right)=\prod_{n=1}^{N}p\left(x_{n}\mid\rho_{n}\right)$
and each $p\left(x_{n}\mid\rho_{n}\right)$ is modeled as a complex
Gaussian prior distribution
\begin{equation}
p\left(x_{n}\mid\rho_{n}\right)=\mathcal{CN}\left(x_{n};0,\rho_{n}^{-1}\right),\forall n=1,...,N.\label{eq:xcondruo}
\end{equation}

\subsection{Structured CS Problem Formulation with a Dynamic Grid\label{sec:CS-Problem-Formulation}}

Recall the CS model with a dynamic grid in the sensing matrix

\begin{equation}
\boldsymbol{y}=\mathbf{A}\left(\boldsymbol{\theta}\right)\boldsymbol{x}+\boldsymbol{w},\label{eq:SCS}
\end{equation}
where ${\color{red}{\color{black}\boldsymbol{w}\in\mathbb{C}^{M\times1}}}$
is the noise vector with independent Gaussian entries $w_{m}\sim\mathcal{CN}\left(w_{m};0,\kappa^{-1}\right)$.
We employ a Gamma distribution with parameters $c$ and $d$ to model
the noise precision, i.e.,
\begin{equation}
p\left(\kappa\right)=\Gamma\left(\kappa;c,d\right).\label{eq:p(gamma)}
\end{equation}
The Gamma distribution can capture the practical distribution of the
noise precision well and is a conjugate of the Gaussian prior. Therefore,
it has been widely used to model the noise precision in Bayesian inference
\cite{LiuAn_CE_Turbo_VBI,LiuAn_directloc_vehicles,OGSBI,Duan_IFSBL,Tzikas_VBI}. 

Given the observation $\boldsymbol{y}$, we aim at computing a Bayesian
estimation of the sparse signal $\boldsymbol{x}$ and support $\boldsymbol{s}$,
i.e., the posterior $p\left(\boldsymbol{x}\mid\boldsymbol{y}\right)$
and $p\left(\boldsymbol{s}\mid\boldsymbol{y}\right)$, and the maximum
likelihood estimation (MLE) of grid parameters $\boldsymbol{\theta}$,
i.e., $\underset{\boldsymbol{\theta}}{\text{argmax}}\ln p\left(\boldsymbol{\theta}\mid\boldsymbol{y}\right)$.
Note that we can also obtain an MMSE estimate of $\boldsymbol{x}$
and a MAP estimate of $\boldsymbol{s}$ from the Bayesian estimation
$p\left(\boldsymbol{x}\mid\boldsymbol{y}\right)$ and $p\left(\boldsymbol{s}\mid\boldsymbol{y}\right)$.

\subsection{Example: Massive MIMO Channel Estimation under Hybrid Beamforming\label{subsec:Example:-Massive-MIMO}}

Consider a narrowband massive MIMO system under a partially connected
hybrid beamforming (PC-HBF) architecture, where the base station (BS)
is equipped with $N_{r}$ antennas and $N_{RF}$ radio frequency (RF)
chains with $N_{RF}<N_{r}$, as illustrated in Fig. \ref{fig:Illustration_BS}.
Each user is equipped with a single antenna. In the channel training
stage, each user transmits an uplink pilot at different pilot symbols
for the BS to estimate the channel. Without loss of generality, we
focus on the channel estimation of one user, the received signal $\boldsymbol{y}\in\mathbb{C}^{N_{RF}\times1}$
at the BS can be written as:
\begin{figure}
\begin{centering}
\includegraphics[width=80mm]{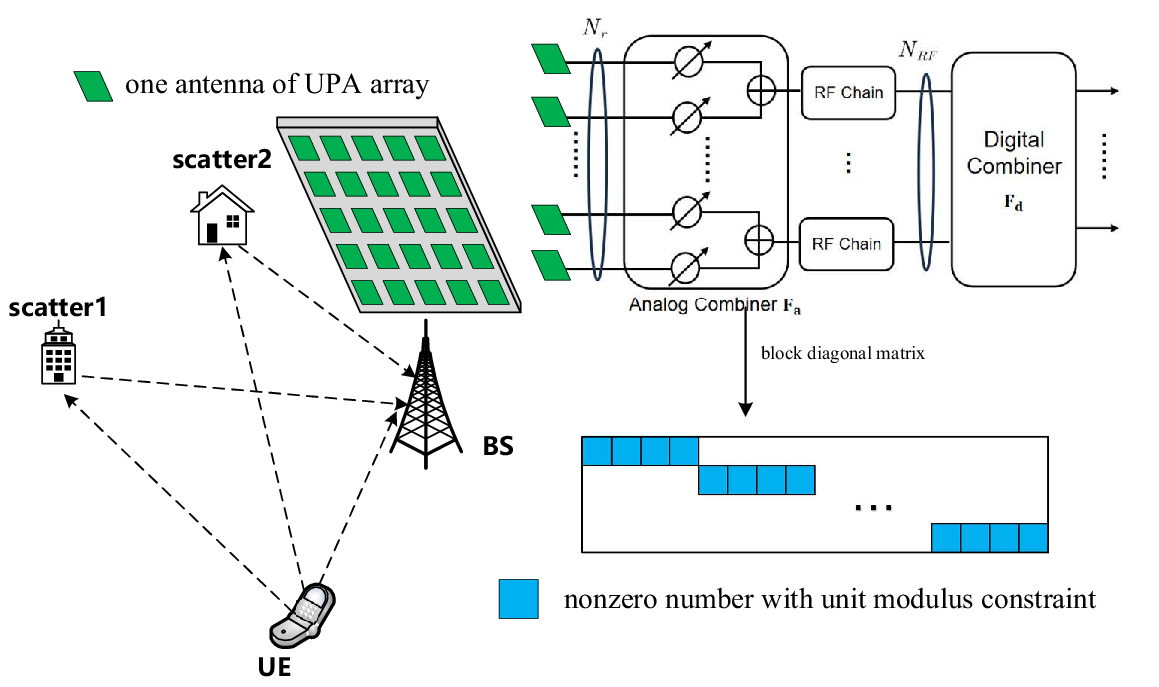}
\par\end{centering}
\caption{\label{fig:Illustration_BS}Illustration of channel estimation in
PC-HBF massive MIMO system.}
\end{figure}

\begin{equation}
\boldsymbol{y}=\mathbf{F}\boldsymbol{h}u+\boldsymbol{w},\label{eq:y_HBF}
\end{equation}
where $u$ is the pilot symbol, $\boldsymbol{h}\in\mathbb{C}^{N_{r}\times1}$
is the channel vector of the user, $\mathbf{F}=\mathbf{F}_{d}\mathbf{F}_{a}\in\mathbb{C}^{N_{RF}\times N_{r}}$,
$\mathbf{F}_{a}\in\mathbb{C}^{N_{RF}\times N_{r}}$ is the analog
combining matrix, $\mathbf{F}_{d}\in\mathbb{C}^{N_{RF}\times N_{RF}}$
is the digital combining matrix, and $\boldsymbol{w}\sim\mathcal{CN}(0,\sigma^{2}\textbf{I})\in\mathbb{C}^{N_{RF}\times1}$
is the additive white Gaussian Noise (AWGN). Since the analog combining
matrix is composed of a series of low-cost phase shifters, the non-zero
elements of analog precoding matrix \textbf{$\mathbf{F}_{a}$} have
unit modulus constraint. Moreover, in PC-HBF architecture, each RF
chain is only connected to a subarray of the antennas, and thus $\mathbf{F}_{a}$
is a block diagonal matrix with each block a row vector of $\frac{N_{r}}{N_{RF}}$
elements, as illustrated in Fig. \ref{fig:Illustration_BS}. In the
simulations, to ensure that $\mathbf{F}$ satisfies the restricted
isometry property (RIP) with high probability, the phases of the non-zero
elements in $\mathbf{F}_{a}$ are chosen to be i.i.d. with a uniform
distribution over $\left[0,2\pi\right]$ and the digital combining
matrix $\mathbf{F}_{d}$ is set as a random orthogonal matrix.

The channel vector $\boldsymbol{h}$ can be represented as:

\begin{equation}
\ensuremath{\boldsymbol{h}=\sum_{k=1}^{K}\alpha_{k}\boldsymbol{a}_{R}\left(\theta_{k},\phi_{k}\right),}
\end{equation}
where $\alpha_{k}$, $\theta_{k}$ and $\phi_{k}$ are the complex
gain, azimuth angle of arrival (AoA), and elevation AoA of the $k$-th
channel path, respectively, and $\boldsymbol{a}_{R}\left(\theta_{k},\phi_{k}\right)$
is the array response vector that depends on the antenna array. In
the simulations, we adopt a uniform planar array (UPA) with $N_{x}$
antennas in the horizon direction and $N_{y}$ antennas in the vertical
direction. The grid-based method is adopted to obtain a sparse representation
of massive MIMO channel for high-accuracy channel estimation. Similar
to \cite{Wanyb_Channel_Extrap}, we introduce a 2D dynamic angular-domain
grid of $Q=N_{1}\times N_{2}$ angle points, where the $q\textrm{-th}$
grid point is denoted by $\left(\theta_{q},\phi_{q}\right)$. Define
$\boldsymbol{\theta}\triangleq\left[\theta_{1},\ldots,\theta_{Q}\right]^{T}$
and $\boldsymbol{\phi}\triangleq\left[\phi_{1},\ldots,\phi_{Q}\right]^{T}$
represent the dynamic grid vectors. The uniform angle grid is usually
chosen as the initial value of the dynamic grid \cite{Xu_SLA_VBI}.

With the definition of the dynamic angular-domain grid, we can reformulate
the received signal model in (\ref{eq:y_HBF}) as:

\begin{equation}
\ensuremath{\boldsymbol{y}=u\mathbf{F}\mathbf{A}\left(\boldsymbol{\theta},\boldsymbol{\phi}\right)\boldsymbol{x}+}\boldsymbol{w},
\end{equation}
with
\begin{equation}
\mathbf{A}\left(\boldsymbol{\theta},\boldsymbol{\phi}\right)\triangleq\left[\boldsymbol{a}\left(\theta_{1},\phi_{1}\right),\ldots,\boldsymbol{a}\left(\theta_{Q},\phi_{Q}\right)\right]\in\mathbb{C}^{N_{r}\times Q},
\end{equation}
$\boldsymbol{x}\in\mathbb{C}^{Q\times1}$ is the angular-domain sparse
channel vector, which has only $K\ll Q$ non-zero elements corresponding
to $K$ paths. Specifically, the $q$-th element of $\boldsymbol{x}$,
denoted by $x_{q}$, is the complex gain of the channel path lying
around the $q$-th angle grid point.

The angular-domain sparse channel vector $\boldsymbol{x}$ exhibits
a 2D clustered sparsity, i.e., if we arrange $\boldsymbol{x}$ into
an $N_{1}\times N_{2}$ matrix, the non-zero elements of $\boldsymbol{x}$
will concentrate on a few non-zero 2D clusters \cite{LiuAn_CE_Burst_LASSO}.
We design a 2D Markov model \cite{Fornasini_2D_MM} to capture the
2D clustered sparsity in the angular domain. For convenience, we use
$s_{n_{1},n_{2}}$ to represent the $\left(\left(n_{2}-1\right)N_{1}+n_{1}\right)\textrm{-th}$
element of $\boldsymbol{s}$. The support vector is modeled as 
\begin{align}
p\left(\boldsymbol{s}\right)= & p\left(s_{1,1}\right)\prod_{n_{1}=1}^{N_{1}}\prod_{n_{2}=2}^{N_{2}}p\left(s_{n_{1},n_{2}}\mid s_{n_{1},n_{2}-1}\right)\nonumber \\
 & \times\prod_{n_{1}=2}^{N_{1}}\prod_{n_{2}=1}^{N_{2}}p\left(s_{n_{1},n_{2}}\mid s_{n_{1}-1,n_{2}}\right),\label{eq:2D_Markov}
\end{align}
with the transition probability given by
\begin{subequations}
\begin{align}
p_{01}^{\textrm{row}} & =p\left(s_{n_{1},n_{2}}=1\mid s_{n_{1},n_{2}-1}=0\right),\\
p_{10}^{\textrm{row}} & =p\left(s_{n_{1},n_{2}}=0\mid s_{n_{1},n_{2}-1}=1\right),\\
p_{01}^{\textrm{col}} & =p\left(s_{n_{1},n_{2}}=1\mid s_{n_{1}-1,n_{2}}=0\right),\\
p_{10}^{\textrm{col}} & =p\left(s_{n_{1},n_{2}}=0\mid s_{n_{1}-1,n_{2}}=1\right).
\end{align}
\end{subequations}
The work in \cite{Fornasini_2D_MM} verified that a set of $\left\{ p_{01}^{\textrm{row}},p_{10}^{\textrm{row}},p_{01}^{\textrm{col}},p_{10}^{\textrm{col}}\right\} $
can be found to satisfy the steady-state condition of the 2D Markov
model, i.e., $p\left(s_{1,1}=1\right)=\lambda$, where $\lambda$
denotes the sparsity level of $\boldsymbol{s}$. As such, the initial
distribution $p\left(s_{1,1}\right)$ is set to be the steady-state
distribution, $p\left(s_{1,1}=1\right)=\lambda$.

\begin{figure}[t]
\begin{centering}
\includegraphics[width=70mm]{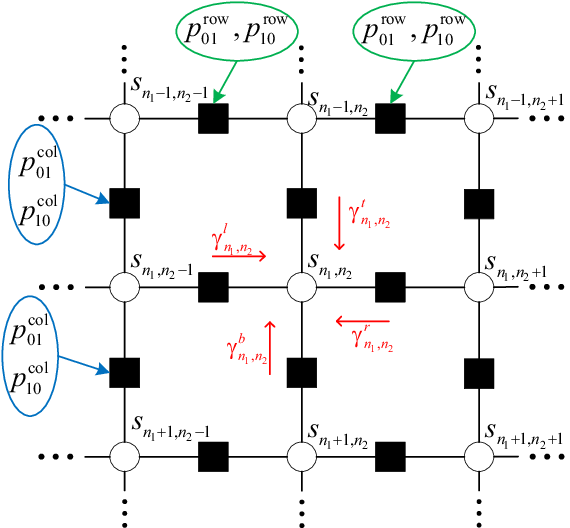}
\par\end{centering}
\caption{\label{fig:factor_2D_Markov}The factor graph of the 2D Markov model.}
\end{figure}

We show the factor graph of the 2D Markov model in Fig. \ref{fig:factor_2D_Markov}.
The value of $\left\{ p_{01}^{\textrm{row}},p_{10}^{\textrm{row}},p_{01}^{\textrm{col}},p_{10}^{\textrm{col}}\right\} $
will affect the structure of clusters. Specifically, smaller $p_{10}^{\textrm{row}}$
and $p_{10}^{\textrm{col}}$ imply a larger average cluster size,
and smaller $p_{01}^{\textrm{row}}$ and $p_{01}^{\textrm{col}}$
imply a larger average gap between two adjacent clusters. 

\section{The proposed Alternating Estimation Framework \label{sec:The-proposed-Alternating} }

It is very challenging to calculate the exact posterior $p\left(\boldsymbol{x}\mid\boldsymbol{y}\right)$
and $p\left(\boldsymbol{s}\mid\boldsymbol{y}\right)$ because the
factor graph of the associated joint probability model has loops.
In this section, we shall propose an alternating estimation (AE) framework
to approximately calculate the marginal posteriors $p\left(\boldsymbol{x}\mid\boldsymbol{y}\right)$
and $p\left(s_{n}\mid\boldsymbol{y}\right),\forall n$, and finds
an approximate solution for MLE problem $\underset{\boldsymbol{\theta}}{\text{argmax}}\ln p\left(\boldsymbol{\theta}\mid\boldsymbol{y}\right)$.
The proposed AE framework is shown in the simulations to achieve a
good performance.

\begin{figure}[t]
\begin{centering}
\includegraphics[width=90mm]{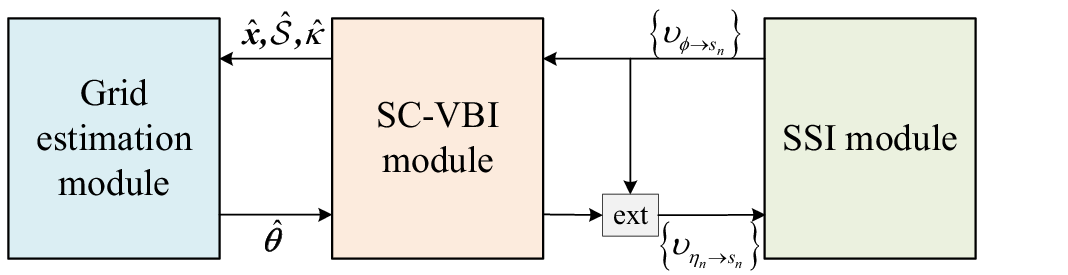}
\par\end{centering}
\caption{\label{fig:AE}The basic modules of the alternating estimation framework.}
\end{figure}

As illustrated in Fig. \ref{fig:AE}, the AE framework iterates among
the following three modules until convergence.
\begin{itemize}
\item \textbf{SC-VBI Module: }For fixed ML estimator $\hat{\boldsymbol{\theta}}$
of $\boldsymbol{\theta}$ from the grid estimation (GE) module and
extrinsic message $\upsilon_{\phi\rightarrow s_{n}}\left(s_{n}\right),\forall n$
from the structured sparse inference (SSI) module (which can be viewed
as prior information for SC-VBI), the SC-VBI module computes a Bayesian
estimation of the collection of parameters $\boldsymbol{v}=\left\{ \boldsymbol{x},\boldsymbol{\rho},\boldsymbol{s},\kappa\right\} $,
from which it outputs MMSE estimators $\hat{\boldsymbol{x}}$ and
$\hat{\kappa}$ of $\boldsymbol{x}$ and $\kappa$, the estimated
support $\hat{\mathcal{S}}$, and extrinsic message $\upsilon_{\eta_{n}\rightarrow s_{n}}\left(s_{n}\right),\forall n$. 
\item \textbf{Grid Estimation Module:} For fixed $\hat{\boldsymbol{x}}$,
$\hat{\mathcal{S}}$, and $\hat{\kappa}$ output from the SC-VBI module,
the grid estimation module refines the dynamic gird by directly maximizing
the likelihood function based on the gradient ascent method with backtracking
line search. 
\item \textbf{Structured Sparse Inference Module:} For fixed extrinsic message
$\upsilon_{\eta_{n}\rightarrow s_{n}}\left(s_{n}\right),\forall n$
from the SC-VBI module (which can be viewed as prior information for
SSI), the SSI module performs sum-product message passing over the
structured sparse prior to exploit the specific sparse structures
in practical applications. 
\end{itemize}

In the following, we shall outline the GE and SSI modules, and explain
the calculation of the extrinsic messages between the SC-VBI and SSI.
The details of the proposed SC-VBI module will be presented in Section
\ref{sec:Turbo-VBI-Algorithm}.

\subsection{The Grid Estimation Module \label{subsec:The-Grid-Estimation}}

To reduce the complexity, we only retain the sparse signals and grid
parameters indexed by the estimated support $\hat{\mathcal{S}}$ and
remove all other signals and grid parameters. In this case, for fixed
$\hat{\boldsymbol{x}}$, $\hat{\mathcal{S}}$ and, $\hat{\kappa}$
output from the SC-VBI module, the logarithmic likelihood function
of $\boldsymbol{\theta}$ is expressed as
\begin{align}
\mathcal{L}\left(\boldsymbol{\theta}_{\hat{\mathcal{S}}}\right) & \triangleq-\hat{\kappa}\left\Vert \boldsymbol{y}-\mathbf{A}_{\hat{\mathcal{S}}}\left(\boldsymbol{\theta}_{\hat{\mathcal{S}}}\right)\hat{\boldsymbol{x}}_{\hat{\mathcal{S}}}\right\Vert ^{2}+C,\label{eq:theta_ML}
\end{align}
where $C$ is a constant, and $\mathbf{A}_{\hat{\mathcal{S}}}\in\mathbb{C}^{M\times\left|\hat{\mathcal{S}}\right|}$
is a submatrix of $\mathbf{A}$ with the column indices lying in $\hat{\mathcal{S}}$.
It is difficult to find the optimal $\boldsymbol{\theta}_{\hat{\mathcal{S}}}$
that maximizes $\mathcal{L}\left(\boldsymbol{\theta}_{\hat{\mathcal{S}}}\right)$
since $\mathcal{L}\left(\boldsymbol{\theta}_{\hat{\mathcal{S}}}\right)$
is non-concave w.r.t. $\boldsymbol{\theta}_{\hat{\mathcal{S}}}$.
In this case, a gradient ascent approach is usually employed to update
$\boldsymbol{\theta}_{\hat{\mathcal{S}}}$. Specifically, in the $i\textrm{-th}$
iteration, the grid parameters are updated as
\begin{align}
\boldsymbol{\theta}_{\hat{\mathcal{S}}}^{\left(i\right)} & =\boldsymbol{\theta}_{\hat{\mathcal{S}}}^{\left(i-1\right)}+\epsilon_{\theta}^{\left(i\right)}\nabla_{\boldsymbol{\theta}_{\hat{\mathcal{S}}}}\mathcal{L}\left(\boldsymbol{\theta}_{\hat{\mathcal{S}}}\right)\mid_{\boldsymbol{\theta}_{\hat{\mathcal{S}}}=\boldsymbol{\theta}_{\hat{\mathcal{S}}}^{\left(i-1\right)}},\label{eq:theta_update}
\end{align}
where $\epsilon_{\theta}^{\left(i\right)}$ is the step size determined
by the Armijo rule. The other grid parameters which are not indexed
by $\hat{\mathcal{S}}$ are set to equal to the previous iteration.
We can apply the above gradient update for $B_{\theta}\geq1$ times,
where $B_{\theta}$ is chosen to achieve a good trade-off between
the per iteration complexity and convergence speed.

\subsection{Calculation of the Extrinsic Messages between SC-VBI and SSI}

\begin{figure}[t]
\begin{centering}
\includegraphics[width=80mm]{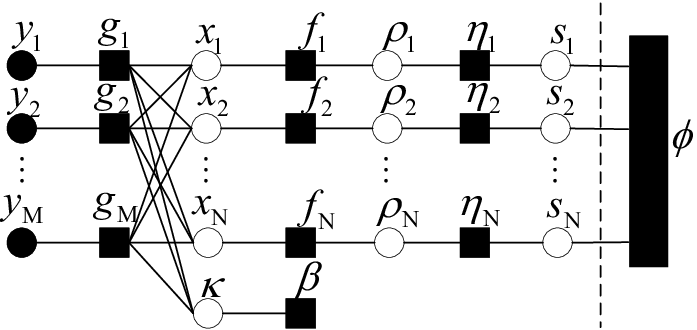}
\par\end{centering}
\caption{\label{fig:factor graph}Factor graph of the joint distribution $p\left(\boldsymbol{v},\boldsymbol{y}\right)$.}
\end{figure}

For fixed grid parameters $\hat{\boldsymbol{\theta}}$, the factor
graph of the joint distribution $p\left(\boldsymbol{v},\boldsymbol{y}\right)$
is shown in Fig. \ref{fig:factor graph}, while the associated factor
nodes are listed in Table \ref{tab:Factors_Table}. Due to the existence
of many loops in this factor graph, we cannot directly apply the message
passing method to calculate the posterior. In this case, the turbo
approach is often used to decouple the observation model and structured
sparse prior model such that the Bayesian inference in each model
is more tractable \cite{Som_TurboAMP}. 
\begin{table}[t]
\caption{\label{tab:Factors_Table}Factors, Distributions and Functional forms
in Fig. \ref{fig:factor graph}. $\mathbf{A}_{m}\left(\boldsymbol{\theta}\right)$
denotes the $m\textrm{-th}$ row of $\mathbf{A}\left(\boldsymbol{\theta}\right)$\textcolor{blue}{.}}

\centering{}%
\begin{tabular}{|c|c|c|}
\hline 
Factor & Distribution & Functional form\tabularnewline
\hline 
\hline 
$g_{m}$ & $p\left(y_{m}\mid\boldsymbol{x},\kappa\right)$ & $\mathcal{CN}\left(y_{m};\mathbf{A}_{m}\left(\boldsymbol{\theta}\right)\boldsymbol{x},\kappa^{-1}\right)$\tabularnewline
\hline 
$f_{n}$ & $p\left(x_{n}\mid\rho_{n}\right)$ & $\mathcal{CN}\left(x_{n};0,\rho_{n}^{-1}\right)$\tabularnewline
\hline 
$\eta_{n}$ & $p\left(\rho_{n}\mid s_{n}\right)$ & $\begin{cases}
\Gamma\left(\rho_{n};a_{n},b_{n}\right), & s_{n}=1\\
\Gamma\left(\rho_{n};\overline{a}_{n},\overline{b}_{n}\right), & s_{n}=0
\end{cases}$\tabularnewline
\hline 
$\phi$ & $p\left(\boldsymbol{s}\right)$ & e.g., (\ref{eq:2D_Markov})\tabularnewline
\hline 
$\beta$ & $p\left(\kappa\right)$ & $\Gamma\left(\kappa;c,d\right)$\tabularnewline
\hline 
\end{tabular}
\end{table}
\begin{figure}[t]
\begin{centering}
\includegraphics[width=80mm]{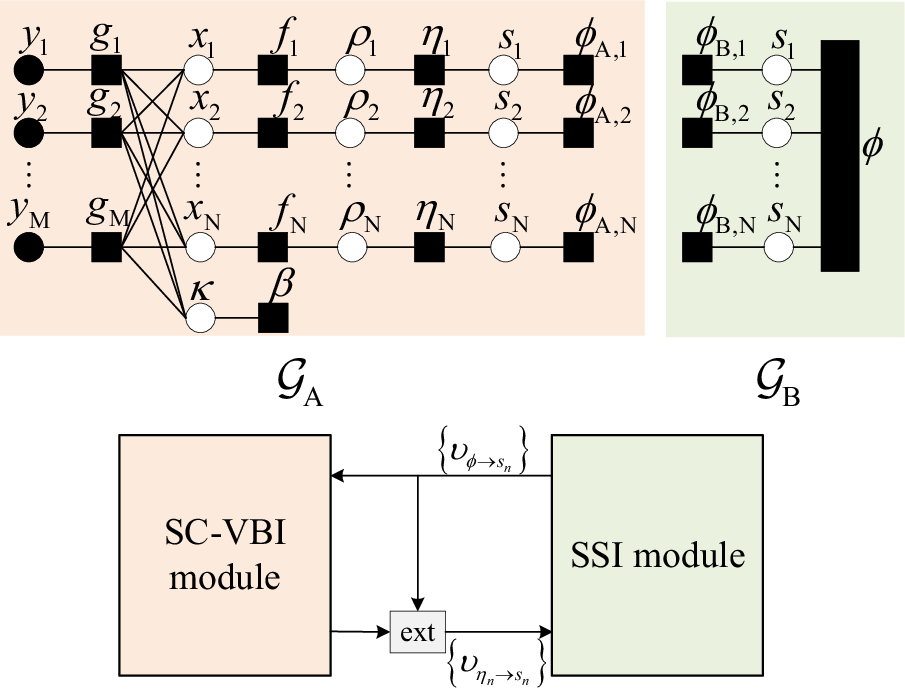}
\par\end{centering}
\caption{\label{fig:Turbo-Algorithm-decoupled-graph}The turbo approach yields
a decoupled factor graph. The extrinsic messages flow between the
SC-VBI module and the SSI module.}
\end{figure}

Specifically, we first partition the factor graph in Fig. \ref{fig:factor graph}
along the dash line into two decoupled subgraphs, denoted by $\mathcal{G}_{\mathrm{A}}$
and $\mathcal{G}_{\mathrm{B}}$, as shown in Fig. \ref{fig:Turbo-Algorithm-decoupled-graph}.
To be more specific, $\mathcal{G}_{\mathrm{A}}$ describes the structure
of the observation model with an independent sparse prior, while $\mathcal{G}_{\mathrm{B}}$
describes the more complicated sparse structure of the support vector
$\boldsymbol{s}$. Then we design SC-VBI and SSI modules to perform
Bayesian inference over the two subgraphs. For $\mathcal{G}_{\mathrm{A}}$,
the SC-VBI module is adopted to compute each variational distribution
approximately. For $\mathcal{G}_{\mathrm{B}}$, sum-product message
passing is adopted to compute the marginal posterior of $\boldsymbol{s}$.
The two modules exchange extrinsic messages until converge. And the
output messages of one module form the priors for another module.
Specifically, the extrinsic messages from SC-VBI to SSI are denoted
by $\left\{ \upsilon_{\eta_{n}\rightarrow s_{n}}\right\} $, while
the extrinsic messages from SSI to SC-VBI are denoted by $\left\{ \upsilon_{\phi\rightarrow s_{n}}\right\} $.
Formally, we define two turbo-iteration factor nodes:
\begin{equation}
\begin{aligned}\phi_{\mathrm{A},n}\left(s_{n}\right) & \triangleq\upsilon_{\phi\rightarrow s_{n}}\left(s_{n}\right),n=1,\ldots,N,\\
\phi_{\mathrm{B},n}\left(s_{n}\right) & \triangleq\upsilon_{\eta_{n}\rightarrow s_{n}}\left(s_{n}\right),n=1,\ldots,N,
\end{aligned}
\label{eq:likelihood_message}
\end{equation}
where $\phi_{\mathrm{A},n}\left(s_{n}\right)$ can be viewed as the
prior information for the SC-VBI Module. And for each turbo iteration,
the extrinsic message $\upsilon_{\eta_{n}\rightarrow s_{n}}\left(s_{n}\right)$
from SC-VBI to SSI can be computed by subtracting the prior information
$\phi_{\mathrm{A},n}\left(s_{n}\right)$ from posterior information,
\begin{equation}
\upsilon_{\eta_{n}\rightarrow s_{n}}\left(s_{n}\right)\varpropto q\left(s_{n}\right)/\phi_{\mathrm{A},n}\left(s_{n}\right),\label{eq:likelihood_moduleA}
\end{equation}
where $q\left(s_{n}\right)$ is the approximate posterior distribution
obtained by the SC-VBI as in (\ref{eq:q(s)}). On the other hand,
the extrinsic message $\upsilon_{\phi\rightarrow s_{n}}\left(s_{n}\right)$
from SSI to SC-VBI can be directly computed by message passing over
$\mathcal{G}_{\mathrm{B}}$.

\subsection{The Structured Sparse Inference Module}

The sum-product message passing algorithm for the SSI module depends
on the specific sparse prior model in practical applications. In this
subsection, we use the massive MIMO channel estimation under the 2D
Markov prior model described in Subsection \ref{subsec:Example:-Massive-MIMO}
as a concrete example to illustrate how the SSI Module performs message
passing.

The factor graph of the MRF model in (\ref{eq:2D_Markov}) is given
in Fig. \ref{fig:factor_2D_Markov}. As discussed in subsection \ref{subsec:Example:-Massive-MIMO},
we use $n_{1},n_{2}$ to index the support vector $\boldsymbol{s}$
in this subsection. To simplify the notation, we use $\pi_{n_{1},n_{2}}^{in}$
to abbreviate $\phi_{\mathrm{B},n}\left(s_{n}=1\right)$ with $n=\left(n_{2}-1\right)N_{1}+n_{1}$.
In the following, we obey the sum-product rule to derive messages
over the factor graph \cite{Ksch_sum-product}.

For $s_{n_{1},n_{2}}$, the input messages from the left, right, top,
and bottom factor nodes, denoted by $\upsilon{}_{n_{1},n_{2}}^{l}$,
$\upsilon_{n_{1},n_{2}}^{r}$, $\upsilon_{n_{1},n_{2}}^{t}$, and
$\upsilon_{n_{1},n_{2}}^{b}$, respectively, follow Bernoulli distributions,
which can be calculated as
\begin{align*}
\upsilon_{n_{1},n_{2}}^{d} & =\gamma_{n_{1},n_{2}}^{d}\delta\left(s_{n_{1},n_{2}}-1\right)+\left(1-\gamma_{n_{1},n_{2}}^{d}\right)\delta\left(s_{n_{1},n_{2}}\right),
\end{align*}
for $d\in\left\{ l,r,t,b\right\} $, with $\gamma_{n_{1},n_{2}}^{d},d\in\left\{ l,r,t,b\right\} $
given at the top of the next page.
\begin{figure*}
\begin{subequations}
\begin{align}
\gamma_{n_{1},n_{2}}^{l}= & \frac{p_{11}^{\textrm{row}}\pi_{n_{1},n_{2}-1}^{in}\prod_{d\in\left\{ l,t,b\right\} }\gamma_{n_{1},n_{2}-1}^{d}+p_{01}^{\textrm{row}}\left(1-\pi_{n_{1},n_{2}-1}^{in}\right)\prod_{d\in\left\{ l,t,b\right\} }\left(1-\gamma_{n_{1},n_{2}-1}^{d}\right)}{\pi_{n_{1},n_{2}-1}^{in}\prod_{d\in\left\{ l,t,b\right\} }\gamma_{n_{1},n_{2}-1}^{d}+\left(1-\pi_{n_{1},n_{2}-1}^{in}\right)\prod_{d\in\left\{ l,t,b\right\} }\left(1-\gamma_{n_{1},n_{2}-1}^{d}\right)},\\
\gamma_{n_{1},n_{2}}^{r}= & \frac{p_{11}^{\textrm{row}}\pi_{n_{1},n_{2}+1}^{in}\prod_{d\in\left\{ r,t,b\right\} }\gamma_{n_{1},n_{2}+1}^{d}+p_{10}^{\textrm{row}}\left(1-\pi_{n_{1},n_{2}+1}^{in}\right)\prod_{n\in\left\{ r,t,b\right\} }\left(1-\gamma_{n_{1},n_{2}+1}^{d}\right)}{\left(p_{11}^{\textrm{row}}+p_{01}^{\textrm{row}}\right)\pi_{n_{1},n_{2}+1}^{in}\prod_{d\in\left\{ r,t,b\right\} }\gamma_{n_{1},n_{2}+1}^{d}+\left(p_{00}^{\textrm{row}}+p_{10}^{\textrm{row}}\right)\left(1-\pi_{n_{1},n_{2}+1}^{in}\right)\prod_{d\in\left\{ r,t,b\right\} }\left(1-\gamma_{n_{1},n_{2}+1}^{d}\right)},\\
\gamma_{n_{1},n_{2}}^{t}= & \frac{p_{11}^{\textrm{col}}\pi_{n_{1}-1,n_{2}}^{in}\prod_{d\in\left\{ l,r,t\right\} }\gamma_{n_{1}-1,n_{2}}^{d}+p_{01}^{\textrm{col}}\left(1-\pi_{n_{1}-1,n_{2}}^{in}\right)\prod_{d\in\left\{ l,r,t\right\} }\left(1-\gamma_{n_{1}-1,n_{2}}^{d}\right)}{\pi_{n_{1}-1,n_{2}}^{in}\prod_{d\in\left\{ l,r,t\right\} }\gamma_{n_{1}-1,n_{2}}^{d}+\left(1-\pi_{n_{1}-1,n_{2}}^{in}\right)\prod_{d\in\left\{ l,r,t\right\} }\left(1-\gamma_{n_{1}-1,n_{2}}^{d}\right)},\\
\gamma_{n_{1},n_{2}}^{b}= & \frac{p_{11}^{\textrm{col}}\pi_{n_{1}+1,n_{2}}^{in}\prod_{d\in\left\{ l,r,b\right\} }\gamma_{n_{1}+1,n_{2}}^{d}+p_{10}^{\textrm{col}}\left(1-\pi_{n_{1}+1,n_{2}}^{in}\right)\prod_{d\in\left\{ l,r,b\right\} }\left(1-\gamma_{n_{1}-1,n_{2}}^{d}\right)}{\left(p_{11}^{\textrm{col}}+p_{01}^{\textrm{col}}\right)\pi_{n_{1}+1,n_{2}}^{in}\prod_{d\in\left\{ l,r,b\right\} }\gamma_{n_{1}-1,n_{2}}^{d}+\left(p_{00}^{\textrm{col}}+p_{10}^{\textrm{col}}\right)\left(1-\pi_{n_{1}+1,n_{2}}^{in}\right)\prod_{d\in\left\{ l,r,b\right\} }\left(1-\gamma_{n_{1}-1,n_{2}}^{d}\right)},
\end{align}
\end{subequations}

\rule[0.5ex]{1\textwidth}{1pt}
\end{figure*}

Then the output message for $s_{n_{1},n_{2}}$ (i.e., the extrinsic
message from the SSI module to the SC-VBI module) can be calculated
as
\begin{align}
\upsilon_{\phi\rightarrow s_{n_{1},n_{2}}} & \propto\prod_{d\in\left\{ l,r,t,b\right\} }\upsilon_{n_{1},n_{2}}^{d}\nonumber \\
 & \propto\pi_{n_{1},n_{2}}^{out}\delta\left(s_{n_{1},n_{2}}-1\right)+\left(1-\pi_{n_{1},n_{2}}^{out}\right)\delta\left(s_{n_{1},n_{2}}\right)
\end{align}
where
\[
\pi_{n}^{out}=\frac{\prod_{d\in\left\{ l,r,t,b\right\} }\gamma_{n_{1},n_{2}}^{d}}{\prod_{d\in\left\{ l,r,t,b\right\} }\gamma_{n_{1},n_{2}}^{d}+\prod_{d\in\left\{ l,r,t,b\right\} }\left(1-\gamma_{n_{1},n_{2}}^{d}\right)}.
\]

\section{Subspace Constrained Variational Bayesian Inference \label{sec:Turbo-VBI-Algorithm}}

\subsection{VBI Problem Formulation for the SC-VBI Module}

Based on the observation $\boldsymbol{y}$ and extrinsic message $\phi_{\mathrm{A},n}\left(s_{n}\right)\triangleq\upsilon_{\phi\rightarrow s_{n}}\left(s_{n}\right),\forall n$
from the SSI module, the SC-VBI module adopts VBI \cite{Tzikas_VBI}
to calculate the approximate marginal posteriors $q\left(\boldsymbol{v}\right)$
for fixed grid parameters $\hat{\boldsymbol{\theta}}$. To be more
specific, the turbo-iteration factor nodes $\phi_{\mathrm{A},n}\left(s_{n}\right),\forall n$
incorporate the prior information from the SSI module. Therefore,
the following prior distribution is assumed when performing the VBI
in the SC-VBI module:
\begin{align}
\hat{p}\left(\boldsymbol{x},\boldsymbol{\rho},\boldsymbol{s}\right) & =\hat{p}\left(\boldsymbol{s}\right)p\left(\boldsymbol{\rho}\mid\boldsymbol{s}\right)p\left(\boldsymbol{x}\mid\boldsymbol{\rho}\right),\nonumber \\
\hat{p}\left(\boldsymbol{s}\right) & =\prod_{n=1}^{N}\phi_{\mathrm{A},n}\left(s_{n}\right).\label{eq:VBIpri}
\end{align}
Since the grid parameter is fixed in SC-VBI, we omit the grid parameter
$\hat{\boldsymbol{\theta}}$ and use $\mathbf{A}$ as a simplified
notation for $\mathbf{A}\left(\hat{\boldsymbol{\theta}}\right)$. 

For convenience, we use $\boldsymbol{v}^{k}$ to denote an individual
variable in $\boldsymbol{v}\triangleq\left\{ \boldsymbol{x},\boldsymbol{\rho},\boldsymbol{s},\kappa\right\} $.
Let $\mathcal{H}=\{k\mid\forall\boldsymbol{v}^{k}\in\boldsymbol{v}\}$
and denote $\hat{p}\left(\boldsymbol{v}\mid\boldsymbol{y}\right)$
as the posterior distribution of $\boldsymbol{v}$ with the prior
$\hat{p}\left(\boldsymbol{x},\boldsymbol{\rho},\boldsymbol{s}\right)$
in (\ref{eq:VBIpri}). Based on the VBI method, the approximate marginal
posterior could be calculated by minimizing the KLD between $\hat{p}\left(\boldsymbol{v}\mid\boldsymbol{y}\right)$
and $q\left(\boldsymbol{v}\right)$, subject to a factorized form
constraint as \cite{Tzikas_VBI}
\begin{eqnarray}
\mathcal{\mathscr{A}}_{\mathrm{VBI}}: & q^{*}\left(\boldsymbol{v}\right) & =\arg\min_{q\left(\boldsymbol{v}\right)}\int q\left(\boldsymbol{v}\right)\ln\frac{q\left(\boldsymbol{v}\right)}{\hat{p}\left(\boldsymbol{v}\mid\boldsymbol{y}\right)}\textrm{d}\boldsymbol{v},\label{eq:KLDmin}\\
\mathrm{s.t.} & q\left(\boldsymbol{v}\right) & =\prod_{k\in\mathcal{H}}q\left(\boldsymbol{v}^{k}\right),\int q\left(\boldsymbol{v}^{k}\right)d\boldsymbol{v}^{k}=1,\label{eq:factorconstrain}\\
 & q\left(\boldsymbol{x}\right) & =\mathcal{CN}\left(\boldsymbol{x};\boldsymbol{\mu},\textrm{diag}\left(\boldsymbol{\sigma}^{2}\right)\right).\label{eq:GScon}
\end{eqnarray}
Note that we add an additional constraint that $q\left(\boldsymbol{x}\right)$
is a Gaussian distribution with mean $\boldsymbol{\mu}$ and diagonal
covariance matrix $\textrm{diag}\left(\boldsymbol{\sigma}^{2}\right)$,
where both the mean $\boldsymbol{\mu}$ and variance vector $\boldsymbol{\sigma}^{2}=\left[\sigma_{1}^{2},...,\sigma_{N}^{2}\right]^{T}$
are optimization variables. This additional constraint is imposed
to avoid the high dimensional matrix inverse for calculating the posterior
covariance matrix. On the other hand, the high dimensional matrix
inverse for calculating the posterior mean will be avoided using the
subspace constrained method. 

Solving $\mathcal{\mathscr{A}}_{\mathrm{VBI}}$ yields a good approximation
of the true posterior $\hat{p}\left(\boldsymbol{v}\mid\boldsymbol{y}\right)$
\cite{Tzikas_VBI}. Since Problem $\mathcal{\mathscr{A}}_{\mathrm{VBI}}$
is non-convex, we aim at finding a stationary solution (denoted by
$q^{*}\left(\boldsymbol{v}\right)$) of $\mathcal{\mathscr{A}}_{\mathrm{VBI}}$,
as defined below.
\begin{defn}
[Stationary Solution]\label{lem:optimality-conditon-1}$q^{*}\left(\boldsymbol{v}\right)=\prod_{k\in\mathcal{H}}q^{*}\left(\boldsymbol{v}^{k}\right)$
is called a stationary solution of Problem $\mathcal{\mathscr{A}}_{\mathrm{VBI}}$
if it satisfies all the constraints in $\mathcal{\mathscr{A}}_{\mathrm{VBI}}$
and $\forall k\in\mathcal{H}$,
\begin{multline*}
q^{*}\left(\boldsymbol{v}^{k}\right)=\\
\arg\min_{q\left(\boldsymbol{v}^{k}\right)}\int\prod_{l\neq k}q^{*}\left(\boldsymbol{v}^{l}\right)q\left(\boldsymbol{v}^{k}\right)\ln\frac{\prod_{l\neq k}q^{*}\left(\boldsymbol{v}^{l}\right)q\left(\boldsymbol{v}^{k}\right)}{\hat{p}\left(\boldsymbol{v}\mid\boldsymbol{y}\right)}.
\end{multline*}
In particular, when $\boldsymbol{v}^{k}=\boldsymbol{x}$, the above
minimization is over the optimization variables $\boldsymbol{\mu},\boldsymbol{\sigma}^{2}$
in $q\left(\boldsymbol{x}\right)=\mathcal{CN}\left(\boldsymbol{x};\boldsymbol{\mu},\textrm{diag}\left(\boldsymbol{\sigma}^{2}\right)\right)$.

\end{defn}
By finding a stationary solution $q^{*}\left(\boldsymbol{v}\right)$
of $\mathcal{\mathscr{A}}_{\mathrm{VBI}}$, we could obtain the approximate
posterior $q^{*}\left(\boldsymbol{v}^{k}\right)\thickapprox p\left(\boldsymbol{v}^{k}\mid\boldsymbol{y}\right)$.

When there is no additional constraint of $q\left(\boldsymbol{x}\right)=\mathcal{CN}\left(\boldsymbol{x};\boldsymbol{\mu},\textrm{diag}\left(\boldsymbol{\sigma}^{2}\right)\right)$,
the problem of finding a stationary solution $q^{*}\left(\boldsymbol{v}\right)$
of $\mathcal{\mathscr{A}}_{\mathrm{VBI}}$ has been addressed in \cite{LiuAn_CE_Turbo_VBI}
using a sparse VBI algorithm, where the update of $q\left(\boldsymbol{x}\right)$
involves a high-dimensional matrix inverse. In the following, we shall
first derive a VBI algorithm for $\mathcal{\mathscr{A}}_{\mathrm{VBI}}$
with constraint $q\left(\boldsymbol{x}\right)=\mathcal{CN}\left(\boldsymbol{x};\boldsymbol{\mu},\textrm{diag}\left(\boldsymbol{\sigma}^{2}\right)\right)$,
which is called independence constrained VBI (IC-VBI) since the approximate
posterior $q\left(\boldsymbol{x}\right)$ are product of independently
Gaussian distributions of $x_{n}$. Then we modify the update equation
of the posterior mean in $q\left(\boldsymbol{x}\right)$ to obtain
a low-complexity SC-VBI algorithm.

\subsection{Derivation of the IC-VBI Algorithm for $\mathcal{\mathscr{A}}_{\mathrm{VBI}}$}

The IC-VBI algorithm finds a stationary solution of $\mathcal{\mathscr{A}}_{\mathrm{VBI}}$
via alternately optimizing each individual density $q\left(\boldsymbol{v}^{k}\right),k\in\mathcal{H}$.
Specifically, for given $q\left(\boldsymbol{v}^{l}\right),\forall l\neq k$
and $\boldsymbol{v}^{k}\neq\boldsymbol{x}$, the optimal $q\left(\boldsymbol{v}^{k}\right)$
that minimizes the KLD in $\mathcal{\mathscr{A}}_{\mathrm{VBI}}$
is given by \cite{Tzikas_VBI}
\begin{equation}
q\left(\boldsymbol{v}^{k}\right)\propto\exp\left(\left\langle \ln\hat{p}\left(\boldsymbol{v},\boldsymbol{y}\right)\right\rangle _{\prod_{l\neq k}q\left(\boldsymbol{v}^{l}\right)}\right),\label{eq:optimal_q}
\end{equation}
where $\left\langle f\left(x\right)\right\rangle _{q(x)}=\int f\left(x\right)q(x)dx$
and $\hat{p}\left(\boldsymbol{v},\boldsymbol{y}\right)$ is the joint
PDF with prior $\hat{p}\left(\boldsymbol{x},\boldsymbol{\rho},\boldsymbol{s}\right)$,
i.e.,
\begin{align}
\hat{p}\left(\boldsymbol{v},\boldsymbol{y}\right) & =p\left(\boldsymbol{y}\mid\boldsymbol{x},\kappa\right)\hat{p}\left(\boldsymbol{x},\boldsymbol{\rho},\boldsymbol{s}\right)p\left(\kappa\right),\label{eq:joint distribution}
\end{align}
where $p\left(\boldsymbol{y}\mid\boldsymbol{x},\kappa\right)=\mathcal{CN}\left(\boldsymbol{y};\mathbf{A}\boldsymbol{x},\kappa^{-1}\mathbf{I}_{M}\right)$
is the likelihood function. However, when $\boldsymbol{v}^{k}=\boldsymbol{x}$,
the optimal $q\left(\boldsymbol{x}\right)$ is no longer given by
(\ref{eq:optimal_q}) due to the independence constraint, and it is
characterized by the following lemma. Note that the expectation $\left\langle f\left(\boldsymbol{v}^{k}\right)\right\rangle _{q\left(\boldsymbol{v}^{k}\right)}$
w.r.t. its own approximate posterior will be simplified as $\left\langle f\left(\boldsymbol{v}^{k}\right)\right\rangle $.
\begin{lem}
\label{lem:ICVBI}For given $q\left(\boldsymbol{\rho}\right),q\left(\boldsymbol{s}\right),q\left(\kappa\right)$,
$\mathcal{\mathscr{A}}_{\mathrm{VBI}}$ is equivalent to the following
optimization problem
\begin{equation}
\underset{\boldsymbol{u},\boldsymbol{\sigma}^{2}}{\text{min}}\boldsymbol{u}^{H}\mathbf{W}\boldsymbol{u}-2\mathfrak{Re}\left\{ \boldsymbol{u}^{H}\boldsymbol{b}\right\} +\sum_{n=1}^{N}\left(W_{n}\sigma_{n}^{2}-\ln\sigma_{n}^{2}\right),\label{eq:EQIC}
\end{equation}
where $\mathbf{W}=\textrm{diag}\left(\left\langle \boldsymbol{\rho}\right\rangle \right)+\left\langle \kappa\right\rangle \mathbf{A}{}^{H}\mathbf{A}$
and $\boldsymbol{b}=\left\langle \kappa\right\rangle \mathbf{A}{}^{H}\boldsymbol{y}$,
$W_{n}$ is the $n$-th diagonal element of $\mathbf{W}$. In other
words, for given $q\left(\boldsymbol{\rho}\right),q\left(\boldsymbol{s}\right),q\left(\kappa\right)$,
the optimal posterior mean $\boldsymbol{\mu}$ and variance vector
$\boldsymbol{\sigma}^{2}$ in $q\left(\boldsymbol{x}\right)$ that
minimizes the KLD subject to $q\left(\boldsymbol{x}\right)=\mathcal{CN}\left(\boldsymbol{x};\boldsymbol{\mu},\textrm{diag}\left(\boldsymbol{\sigma}^{2}\right)\right)$
are given by the 
\begin{eqnarray*}
\boldsymbol{\mu} & = & \underset{\boldsymbol{u}}{\text{argmin}}\varphi\left(\boldsymbol{u}\right)\triangleq\boldsymbol{u}^{H}\mathbf{W}\boldsymbol{u}-2\mathfrak{Re}\left\{ \boldsymbol{u}^{H}\boldsymbol{b}\right\} ,\\
\boldsymbol{\sigma}^{2} & = & \left[W_{1}^{-1},W_{2}^{-1},...,W_{N}^{-1}\right].
\end{eqnarray*}
\end{lem}

Please refer to Appendix \ref{subsec:Proof-of-lemmaic} for the detailed
proof. Based on \eqref{eq:optimal_q} and Lemma \ref{lem:ICVBI},
the update equations of all variables are given in the subsequent
subsections.

\subsubsection{Update Equation for $q\left(\boldsymbol{x}\right)$}

From Lemma \ref{lem:ICVBI}, $q\left(\boldsymbol{x}\right)$ can be
derived as
\begin{equation}
q(\boldsymbol{x})=\mathcal{CN}\left(\mathbf{x};\boldsymbol{\mu},\boldsymbol{\sigma}^{2}\right),\label{eq:poster_x}
\end{equation}
where $\boldsymbol{\mu}=\mathbf{W}^{-1}\boldsymbol{b}$ and $\boldsymbol{\sigma}^{2}=\left[W_{1}^{-1},W_{2}^{-1},...,W_{N}^{-1}\right].$ 

\subsubsection{Update Equation for $q\left(\boldsymbol{\rho}\right)$}

$q\left(\boldsymbol{\rho}\right)$ can be derived as
\begin{equation}
q\left(\boldsymbol{\rho}\right)=\prod_{n=1}^{N}\Gamma\left(\rho_{n};\widetilde{a}_{n},\widetilde{b}_{n}\right),\label{eq:poster_rho}
\end{equation}
where the approximate posterior parameters are given by:
\begin{align}
\widetilde{a}_{n}= & \left\langle s_{n}\right\rangle a_{n}+\left\langle 1-s_{n}\right\rangle \overline{a}_{n}+1,\label{eq:ab_tilde}\\
\widetilde{b}_{n}= & \left\langle s_{n}\right\rangle b_{n}+\left\langle 1-s_{n}\right\rangle \overline{b}_{n}+\left\langle \left|x_{n}\right|^{2}\right\rangle .\nonumber 
\end{align}

\subsubsection{Update Equation for $q\left(\boldsymbol{s}\right)$}

$q\left(\boldsymbol{s}\right)$ can be derived as 
\begin{equation}
q\left(\boldsymbol{s}\right)=\prod_{n=1}^{N}\left(\widetilde{\lambda}_{n}\right)^{s_{n}}\left(1-\widetilde{\lambda}_{n}\right)^{1-s_{n}},\label{eq:q(s)}
\end{equation}
where $\widetilde{\lambda}_{n}$ is given by
\begin{equation}
\widetilde{\lambda}_{n}=\frac{\lambda_{n}C_{n}}{\lambda_{n}C_{n}+\left(1-\lambda_{n}\right)\overline{C}_{n}},\label{eq:lambda_n_tilde}
\end{equation}
with $\lambda_{n}=\phi_{\mathrm{A},n}\left(s_{n}=1\right)$, $C_{n}=\dfrac{b_{n}^{a_{n}}}{\Gamma\left(a_{n}\right)}\exp\left(\left(a_{n}-1\right)\left\langle \ln\rho_{n}\right\rangle -b_{n}\left\langle \rho_{n}\right\rangle \right)$,
and $\overline{C}_{n}=\dfrac{\overline{b}_{n}^{\overline{a}_{n}}}{\Gamma\left(\overline{a}_{n}\right)}\exp\left(\left(\overline{a}_{n}-1\right)\left\langle \ln\rho_{n}\right\rangle -\overline{b}_{n}\left\langle \rho_{n}\right\rangle \right)$.
Here, $\Gamma\left(\cdot\right)$ denotes the gamma function.

\subsubsection{Update Equation for $q\left(\kappa\right)$}

The posterior distribution $q\left(\kappa\right)$ is given by
\begin{equation}
q\left(\kappa\right)=\Gamma\left(\kappa;\widetilde{c},\widetilde{d}\right),\label{eq:q(gamma)}
\end{equation}
where the parameters $\widetilde{c}$ and $\widetilde{d}$ are given
by
\begin{align}
\widetilde{c} & =c+M,\label{eq:gamma_post_old}\\
\widetilde{d} & =d+\left\langle \left\Vert \boldsymbol{y}-\mathbf{A}\boldsymbol{x}\right\Vert ^{2}\right\rangle _{q\left(\boldsymbol{x}\right)}.\nonumber 
\end{align}
The expectations used in the above update equations are summarized
as follows:
\[
\begin{aligned}\left\langle \rho_{n}\right\rangle = & \begin{aligned}\dfrac{\widetilde{a}_{n}}{\widetilde{b}_{n}} &  & \left\langle \boldsymbol{\rho}\right\rangle = & \left[\bigl\langle\rho_{1}\bigr\rangle,\ldots,\bigl\langle\rho_{N}\bigr\rangle\right]^{T} & \left\langle s_{n}\right\rangle  & =\widetilde{\lambda}_{n}\end{aligned}
\\
\left\langle \gamma\right\rangle = & \begin{aligned}\dfrac{\widetilde{c}}{\widetilde{d}} &  & \left\langle x_{n}^{2}\right\rangle = & \left|\mu_{n}\right|^{2}+\sigma_{n}^{2} & \left\langle \ln\rho_{n}\right\rangle = & \psi\left(\widetilde{a}_{n}\right)-\ln\widetilde{b}_{n},\end{aligned}
\end{aligned}
\]
where $\mu_{n}$ is the $n\textrm{-th}$ element of $\boldsymbol{\mu}$
and $\psi\left(\cdot\right)\triangleq d\ln\left(\Gamma\left(\cdot\right)\right)$
denotes the logarithmic derivative of the gamma function.

\subsection{The Proposed SC-VBI Algorithm}

\subsubsection{The Basic Version}

The complexity of the IC-VBI algorithm in the above is dominated by
the high-dimensional matrix inverse operation $\boldsymbol{\mu}=\mathbf{W}^{-1}\boldsymbol{b}$
in the update of $q\left(\boldsymbol{x}\right)$. Recall that posterior
mean $\boldsymbol{\mu}$ is the optimal solution of the following
quadratic programming problem: 
\begin{equation}
\underset{\boldsymbol{u}}{\text{min}}\varphi\left(\boldsymbol{u}\right)\triangleq\boldsymbol{u}^{H}\mathbf{W}\boldsymbol{u}-2\mathfrak{Re}\left\{ \boldsymbol{u}^{H}\boldsymbol{b}\right\} .\label{eq:QP}
\end{equation}
Let $\hat{\mathcal{S}}$ denote the estimated support of $\boldsymbol{x}$
obtained in the previous iteration, as will be detailed soon. If $\hat{\mathcal{S}}=\mathcal{S}$,
then the LMMSE of $\boldsymbol{x}_{\hat{\mathcal{S}}}$, given by
\begin{equation}
\boldsymbol{\mu}_{\hat{\mathcal{S}}}^{0}=\left\langle \kappa\right\rangle \mathbf{W}_{\hat{\mathcal{S}}}^{-1}\mathbf{A}_{\hat{\mathcal{S}}}^{H}\boldsymbol{y},\label{eq:mu0}
\end{equation}
should be a very good estimation for the non-zero elements in $\boldsymbol{x}$,
where $\mathbf{W}_{\hat{\mathcal{S}}}\in\mathbb{C}^{\left|\hat{\mathcal{S}}\right|\times\left|\hat{\mathcal{S}}\right|}$
is a submatrix of $\mathbf{W}$ with the column/row indices lying
in $\hat{\mathcal{S}}$.

In other words, if we define a vector $\boldsymbol{\mu}^{0}$ such
that $\boldsymbol{\mu}_{\hat{\mathcal{S}}}^{0}=\left\langle \kappa\right\rangle \mathbf{W}_{\hat{\mathcal{S}}}^{-1}\mathbf{A}_{\hat{\mathcal{S}}}^{H}\boldsymbol{y}$
and $\boldsymbol{\mu}_{n}^{0}=\boldsymbol{0},\forall n\notin\hat{\mathcal{S}}$,
then $\boldsymbol{\mu}^{0}$ is expected to provide a good initial
solution for Problem (\eqref{eq:QP}). Then starting from this initial
point, we apply the gradient update for $B_{x}\geq1$ times. Specifically,
in the $i\textrm{-th}$ iteration, the posterior mean is updated as
\begin{align}
\boldsymbol{\mu}^{\left(i\right)} & =\boldsymbol{\mu}^{\left(i-1\right)}-\epsilon_{x}^{\left(i\right)}\nabla_{\boldsymbol{u}}\varphi\left(\boldsymbol{u}\right)\mid_{\boldsymbol{u}=\boldsymbol{\mu}^{\left(i-1\right)}},\label{eq:Gradx}
\end{align}
where $\epsilon_{x}^{\left(i\right)}$ is the step size determined
by the Armijo rule, 
\begin{equation}
\nabla_{\boldsymbol{u}}\varphi\left(\boldsymbol{u}=\boldsymbol{\mu}^{\left(i-1\right)}\right)=\mathbf{W}\boldsymbol{\mu}^{\left(i-1\right)}-\left\langle \kappa\right\rangle \mathbf{A}^{H}\boldsymbol{y},
\end{equation}
and $B_{x}$ is chosen to achieve a good tradeoff between the per
iteration complexity and convergence speed.

\subsubsection{Support Estimation}

Once we obtain the estimated posterior mean $\boldsymbol{\mu}$, we
can obtain an estimated support $\hat{\mathcal{S}}$ for the next
iteration by finding the elements in $\boldsymbol{\mu}$ with sufficiently
large energy. Specifically, we set a small threshold $\varepsilon>0$
and compare each element of $\boldsymbol{\mu}$ with the threshold
to obtain the estimated support as
\begin{equation}
\hat{\mathcal{S}}\triangleq\left\{ n\mid\forall\left|\boldsymbol{\mu}_{n}\right|^{2}>\varepsilon\right\} ,
\end{equation}
where the threshold $\varepsilon$ is chosen according to the noise
power $1/\left\langle \kappa\right\rangle $. A good choice for $\varepsilon$
is 2 to 3 times the noise power $1/\left\langle \kappa\right\rangle $.
Another option is to choose the largest $\left|\hat{\mathcal{S}}\right|$
elements that contain most of the total energy of $\boldsymbol{\mu}$,
e.g., contain 95\% of the total energy.

In the first iteration, we can simply set $\hat{\mathcal{S}}$ according
to the available prior information in practice (e.g., in massive MIMO
channel estimation, we may obtain a rough estimate of the support
according to the previously estimated channels) or a simple baseline
algorithm (e.g., using OMP).

\subsubsection{Robust Design Against Support Estimation Error\label{subsec:Robust-Design-Against}}

To improve the robustness of the algorithm against the support estimation
error, we choose the best initial point from the following two choices:
1) $\boldsymbol{\mu}_{\hat{\mathcal{S}}}^{0}$ obtained by the subspace
constrained matrix inverse as in (\ref{eq:mu0}); 2) the posterior
mean from the previous iteration, depending on which choice gives
a lower value of the objective function $\varphi\left(\boldsymbol{\mu}^{0}\right)$.
The motivations behind those two choices are as follows. In the first
few iterations, $\boldsymbol{\mu}_{\hat{\mathcal{S}}}^{0}$ is expected
to be a good choice because the subspace constrained matrix inverse
can be used to accelerate the initial convergence speed, compared
to the MM-based methods in \cite{Duan_IFSBL,Xu_Turbo-IFVBI,Xu_SLA_VBI}
which completely avoid any matrix inverse. As the iterations go on,
the posterior mean from the previous iteration becomes more and more
accurate and it may also become a good initial point for the gradient
update in \eqref{eq:Gradx}.

\subsection{Convergence Analysis}

The convergence analysis for IC-VBI is straightforward. Each step
of IC-VBI monotonically decreases the KLD until convergence to a stationary
point. However, the convergence analysis for SC-VBI is quite challenging.
First, the support estimation error may cause the algorithm fluctuate.
Fortunately, this problem can be addressed by the robust design mentioned
above, which ensures that the KLD is at least non-increasing. Second,
when updating $q\left(\boldsymbol{x}\right)$, we do not solve the
subproblem in \eqref{eq:EQIC} exactly but only apply a few gradient
update for the posterior mean $\boldsymbol{\mu}$ with the initial
point give by the robust design described above. In the following,
we show that the proposed SC-VBI still converges to a stationary point
of $\mathcal{\mathscr{A}}_{\mathrm{VBI}}$, despite the above approximations
and inexact update.
\begin{thm}
[Convergence of SC-VBI]\label{thm:Convergence-of-SC-VBI}With the
robust design in Subsection \eqref{subsec:Robust-Design-Against},
the SC-VBI algorithm monotonically decrease the KLD objective in $\mathcal{\mathscr{A}}_{\mathrm{VBI}}$,
and every limiting point $q^{*}\left(\boldsymbol{v}\right)=\prod_{k\in\mathcal{H}}q^{*}\left(\boldsymbol{v}^{k}\right)$
generated by the SC-VBI is a stationary solution of Problem $\mathcal{\mathscr{A}}_{\mathrm{VBI}}$.
\end{thm}

Please refer to Appendix \ref{subsec:Proof-of-TheoremSCVBI} for the
detailed proof. The overall AE framework with SC-VBI as the key module
(AE-SC-VBI) is summarized in Algorithm \ref{AE-SC-VBI}. In the first
iteration, we may run the SC-VBI module for multiple times to obtain
a better initial estimation, and in the rest iterations, we only need
to run the SC-VBI module once per iteration to make the overall convergence
speed faster.

\begin{algorithm}[t]
\begin{singlespace}
{\small{}\caption{\label{AE-SC-VBI}AE-SC-VBI algorithm}
}{\small\par}

\textbf{Input:} $\boldsymbol{y}$, $\mathbf{A}\left(\boldsymbol{\theta}\right)$,
maximum iteration number $I$.

\textbf{Output:} $\hat{\boldsymbol{x}}$ and $\hat{\boldsymbol{\theta}}$.

\begin{algorithmic}[1]

\FOR{${\color{blue}{\color{black}k=1,\cdots,I}}$}

\STATE \textbf{SC-VBI Module:}

\STATE Initialize the distribution functions $q\left(\boldsymbol{s}\right)$,
$q\left(\boldsymbol{\rho}\right)$ and $q\left(\kappa\right)$.

\STATE Update $q^{k}\left(\boldsymbol{x}\right)$ using (\ref{eq:poster_x}),
where the posterior mean $\boldsymbol{\mu}$ is obtained by performing
the gradient update (\ref{eq:Gradx}) for $B_{x}$ times with initial
point in (\ref{eq:mu0}).

\STATE Update $q^{k}\left(\boldsymbol{\rho}\right)$ using (\ref{eq:poster_rho}).

\STATE Update $q^{k}\left(\boldsymbol{s}\right)$ using (\ref{eq:q(s)}).

\STATE Update $q^{k}\left(\kappa\right)$ using (\ref{eq:q(gamma)}).

\STATE Obtain the MMSE estimators $\hat{\boldsymbol{x}}$ and $\hat{\kappa}$
of $\boldsymbol{x}$ and $\kappa$ as the posterior mean of $q^{k}\left(\boldsymbol{x}\right)$
and $q^{k}\left(\kappa\right)$, and calculate the estimated support
$\hat{\mathcal{S}}$ from $\hat{\boldsymbol{x}}$.

\STATE Calculate the extrinsic information based on (\ref{eq:likelihood_moduleA}),
send $\upsilon_{\eta_{n}\rightarrow s_{n}}\left(s_{n}\right)$ to
SSI Module.

\STATE\textbf{ Structured Sparse Inference(SSI) Module:}

\STATE Perform message passing over the support subgraph $\mathcal{G}_{\mathrm{B}}$,
and send $\upsilon_{\phi\rightarrow s_{n}}\left(s_{n}\right)$ to
SC-VBI Module to update the estimated support $\hat{\mathcal{S}}$.

\STATE \textbf{Grid Estimation(GE) Module:}

\STATE Given fixed $\hat{\boldsymbol{x}}$, $\hat{\mathcal{S}}$
and, $\hat{\kappa}$, construct the logarithmic likelihood function
of $\boldsymbol{\theta}$ using (\ref{eq:theta_ML}).

\STATE Obtain the ML estimator $\hat{\boldsymbol{\theta}}$ of $\boldsymbol{\theta}$
by performing the gradient update (\ref{eq:theta_update}) for $B_{\theta}$
times.

\ENDFOR

\STATE Output $\hat{\boldsymbol{x}}$ and $\hat{\boldsymbol{\theta}}$.

\end{algorithmic}
\end{singlespace}
\end{algorithm}

\subsection{Comparison with Existing Algorithms}

We compare the proposed SC-VBI with three stare-of-the-art CS-based
algorithms, Turbo-CS \cite{Yuan_TurboCS}, Turbo-VBI \cite{LiuAn_CE_Turbo_VBI},
and IF-VBI \cite{Xu_Turbo-IFVBI}. Both Turbo-CS and Turbo-VBI involve
a matrix inverse each iteration, whose complexity is $\mathcal{O}\left(N^{3}\right)$
per iteration. Although the IF-VBI algorithm has completely avoided
any matrix inverse by using the MM method, an additional SVD operation
is added. In the proposed SC-VBI, we constrain the matrix inverse
in the subspace of the estimated support, which helps to accelerate
the convergence speed and improves the convergence accuracy. Moreover,
the gradient update in \eqref{eq:Gradx} in the SC-VBI uses backtracking
line search to find a better step size.

Table \ref{tab:Factors_Table} compares the complexity order of different
CS-based algorithms. The complexity order of the proposed SC-VBI is
among the lowest, i.e., $\mathcal{O}\left(I\left(NM+\left|\hat{\mathcal{S}}\right|^{3}\right)\right)$.
Due to the aforementioned advantages of SC-VBI, simulations show that
SC-VBI can achieve a much better tradeoff between complexity per iteration,
convergence speed, and performance compared to state-of-the-art baselines.
\begin{table}[t]
\caption{\label{tab:Complexity}Complexity order of different algorithms, where
$I$ represents the iteration number.}

\centering{}%
\begin{tabular}{|c|c|}
\hline 
\multicolumn{1}{|c|}{Algorithms} & \multicolumn{1}{c|}{Complexity order}\tabularnewline
\hline 
Turbo-CS & $\mathcal{O}\left(IN^{3}+N^{2}M\right)$\tabularnewline
\hline 
VBI & $\mathcal{O}\left(IN^{3}+N^{2}M\right)$\tabularnewline
\hline 
IF-VBI & $\mathcal{O}\left(N^{3}+INM\right)$\tabularnewline
\hline 
SC-VBI & $\mathcal{O}\left(I\left(NM+\left|\hat{\mathcal{S}}\right|^{3}\right)\right)$\tabularnewline
\hline 
\end{tabular}
\end{table}

\section{Simulations \label{sec:Applications}}

In this section, we use the massive MIMO channel estimation problem
described in Section \ref{sec:System-Model} as an example to demonstrate
the advantages of the proposed AE-SC-VBI algorithm. The baseline algorithms
considered in the simulations are described below.
\begin{itemize}
\item \textbf{EM-based Turbo-VBI (EM-Turbo-VBI) \cite{LiuAn_CE_Turbo_VBI,LiuAn_directloc_vehicles}:}
The E-step is the Turbo-VBI algorithm and the M-step performs gradient
ascent update.
\item \textbf{EM-based Turbo-IF-VBI (EM-Turbo-IF-VBI)} \textbf{\cite{Duan_IFSBL}:}
The E-step is the low complexity Turbo-IF-VBI algorithm without matrix
inverse, and the M-step performs gradient ascent update.
\item \textbf{EM-based Turbo-CS (EM-Turbo-CS) \cite{Yuan_TurboCS,LiuAn_TurboOAMP}:}
The E-step is the Turbo-CS algorithm, and the M-step performs gradient
ascent update.
\end{itemize}

\subsection{Implementation Details}

In the simulations, the BS is equipped with a UPA consisting of $N_{x}=32$
horizontal antennas and $N_{y}=72$ vertical antennas, for a total
of $N_{r}=2304$ antennas. Define the compression ratio as $\frac{N_{r}}{N_{RF}}$.
The number of azimuth AoA grid points is set to $N_{1}=32$ and the
number of elevation AoA grid points is set to $N_{2}=18$, since the
elevation angle has a prior range between $-\frac{1}{6}\pi$ and $0$
in practical systems. The pilot symbol $u$ is generated with a random
phase under unit power constraint. We consider a narrowband multi-path
channel generated by 3GPP TR 38.901 model \cite{3GPP_channel_model}.
The normalized mean square error (NMSE) is used as the performance
metric for channel estimation. All the algorithms use the 2D-Markov
structured prior model without special announcement.\textcolor{blue}{}
\begin{figure}
\begin{centering}
\textcolor{blue}{\includegraphics[width=80mm]{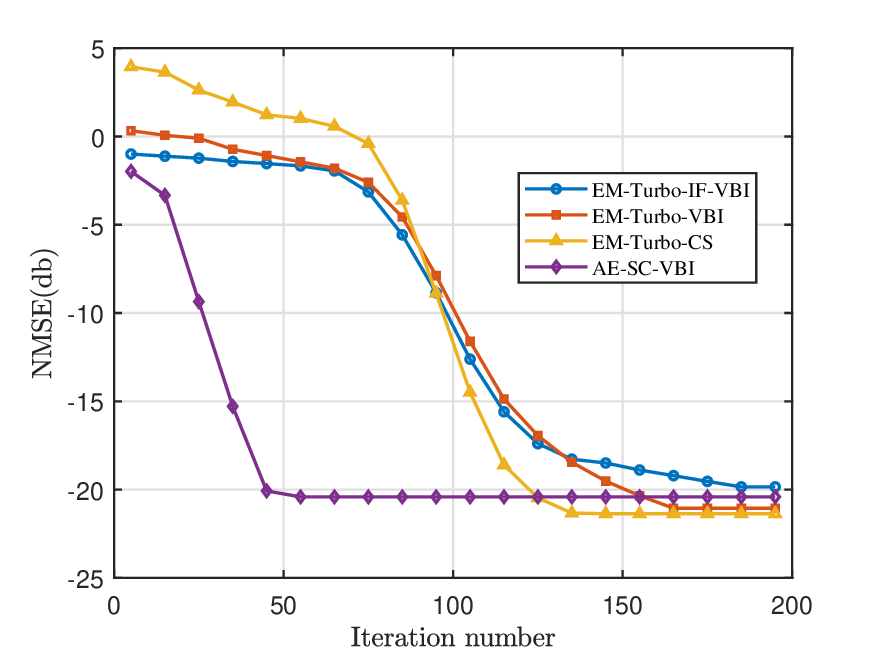}}
\par\end{centering}
\textcolor{blue}{\caption{\label{fig:with_iter}NMSE of channel estimation versus the iteration
number.}
}
\end{figure}

\subsection{Convergence Behavior}

In Fig. \ref{fig:with_iter}, we compare the convergence behavior
of different methods in terms of channel estimation NMSE, we set $\textrm{SNR}=10\ \textrm{dB}$
and the compression ratio is 4. Since the AE-SC-VBI algorithm updates
all variables at the same timescale in an alternative way, while the
other three baseline algorithms update the dynamic grid at a slower
timescale, the AE-SC-VBI converges significantly faster than the other
algorithms. Moreover, even for fixed grid, AE-SC-VBI still converges
much faster than EM-Turbo-IF-VBI because of the introduction of the
subspace constrained matrix inverse. After convergence, the steady-state
performance of each algorithms is similar, which indicates that the
AE-SC-VBI can accelerate the convergence speed while achieving comparable
performance with state-of-the-art methods.\textcolor{blue}{}
\begin{figure}
\begin{centering}
\textcolor{blue}{\includegraphics[width=80mm]{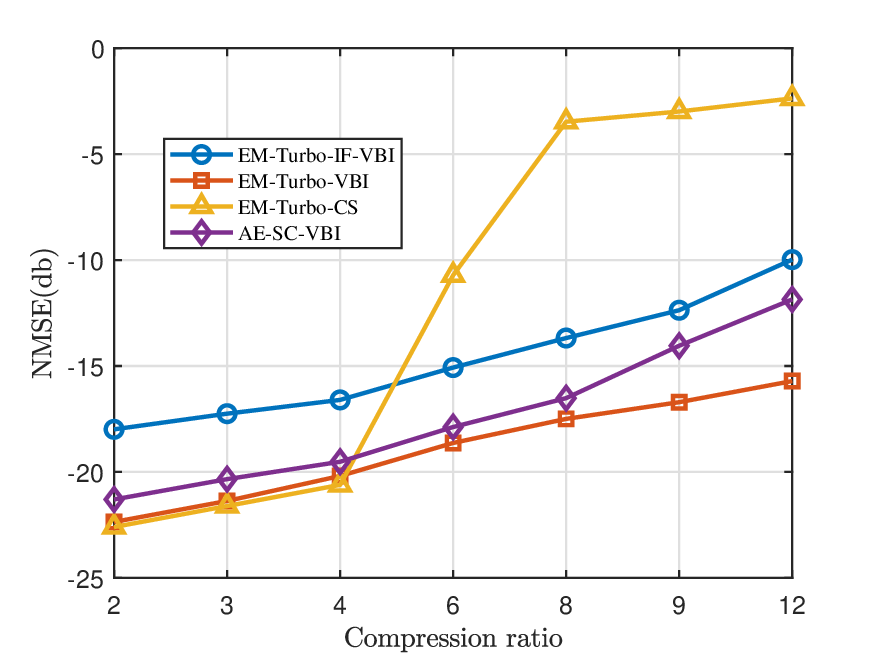}}
\par\end{centering}
\textcolor{blue}{\caption{{\small{}\label{fig:NMSE_with_CR}NMSE of channel estimation versus
the }compress ratio}
}
\end{figure}

\subsection{Influence of Number of RF Chains}

\begin{figure}
\begin{centering}
\textcolor{blue}{\includegraphics[width=80mm]{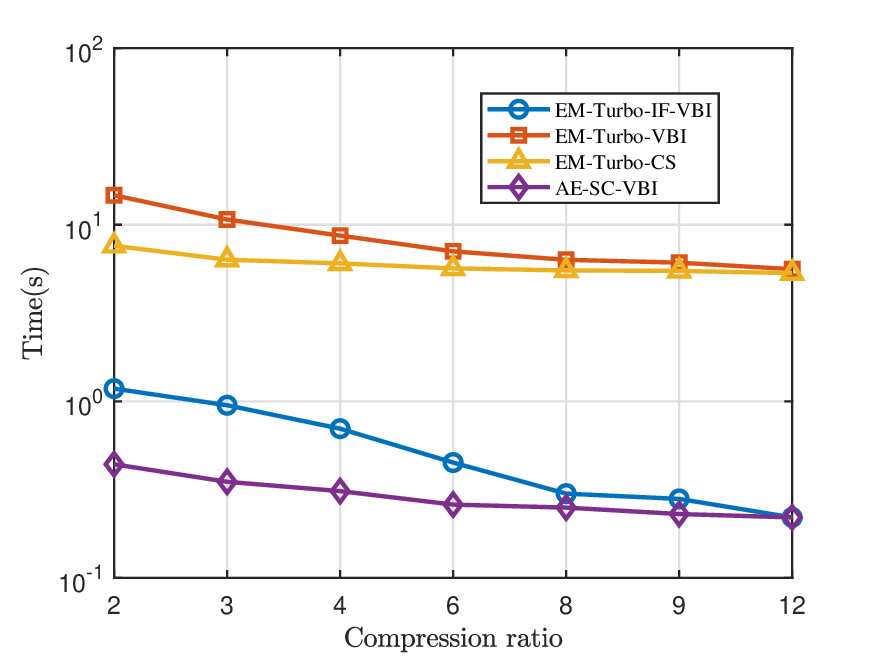}}
\par\end{centering}
\textcolor{blue}{\caption{\label{fig:CPU-time}CPU time versus the compress ratio}
}
\end{figure}

In Fig. \ref{fig:NMSE_with_CR}, we compare the NMSE performance versus
compression ratios when $\textrm{SNR}=10\ \textrm{dB}$. To show that
AE-SC-VBI can achieve a better performance than EM\textbf{-}Turbo-IF-VBI
and meanwhile consumes less computing power, the total iteration time
of EM-Turbo-IF-VBI is set to be slightly greater than AE-SC-VBI. In
Fig. \ref{fig:CPU-time}, we measure the CPU time of each algorithm
in different compression ratios via MATLAB on a laptop computer with
a 2.5 GHz CPU.

It can be observed that the the CPU time of EM-Turbo-CS and EM-Turbo-VBI
is significantly higher than that of EM-Turbo-IF-VBI and AE-SC-VBI,
and the CPU time of AE-SC-VBI is significantly lower than that of
EM-Turbo-IF-VBI when the compression ratio is low, because the SVD
in EM-Turbo-IF-VBI has a higher complexity at low compression ratios.
Besides, although the EM-Turbo-CS algorithm has a good performance
at low compression ratios, its performance significantly degrades
when the compression ratio exceeds 4. As a comparison, the three VBI
algorithms can still achieve good performance at high compression
ratios. The proposed AE-SC-VBI algorithm outperforms the EM-Turbo-IF-VBI
algorithm at all compression ratios and meanwhile consumes less CPU
time. In fact, the performance of AE-SC-VBI algorithm can even approach
that of EM-Turbo-VBI who has a much higher complexity order.\textcolor{blue}{}

\subsection{Influence of SNR}

\begin{figure}
\begin{centering}
\textcolor{blue}{\includegraphics[width=80mm]{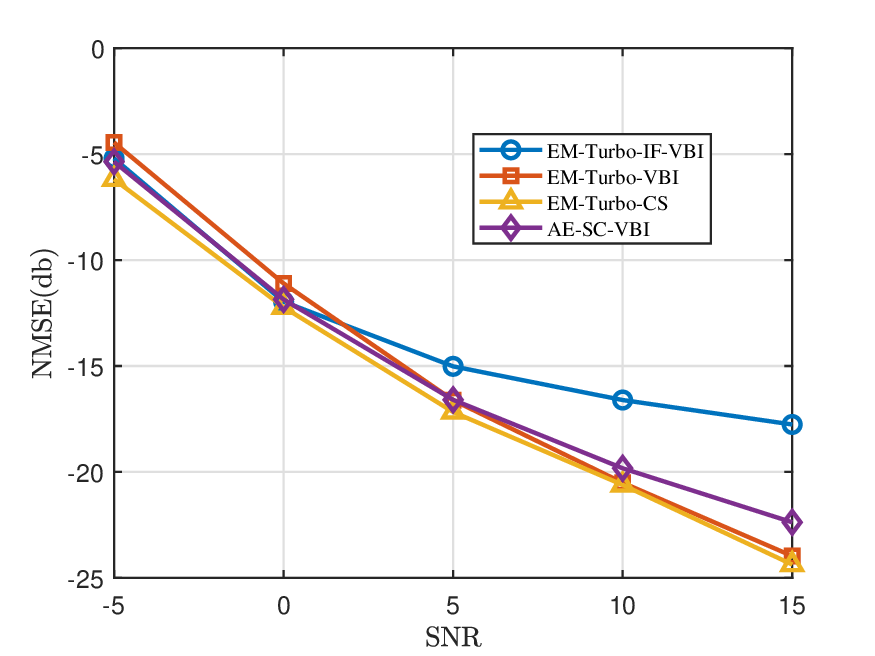}}
\par\end{centering}
\textcolor{blue}{\caption{{\small{}\label{fig:SNR}NMSE of channel estimation versus the SNR}}
}
\end{figure}

In Fig. \ref{fig:SNR}, we compare the NMSE performance versus SNR,
where the compression ratio is set to 4. Our proposed AE-SC-VBI algorithm
significantly outperforms the EM-Turbo-IF-VBI algorithm in the high
SNR region and has similar performance compared to the high-complexity
EM-based Turbo-VBI/Turbo-CS algorithms.\textcolor{blue}{}

\subsection{Influence of Number of Channel Paths}

\begin{figure}
\begin{centering}
\textcolor{blue}{\includegraphics[width=80mm]{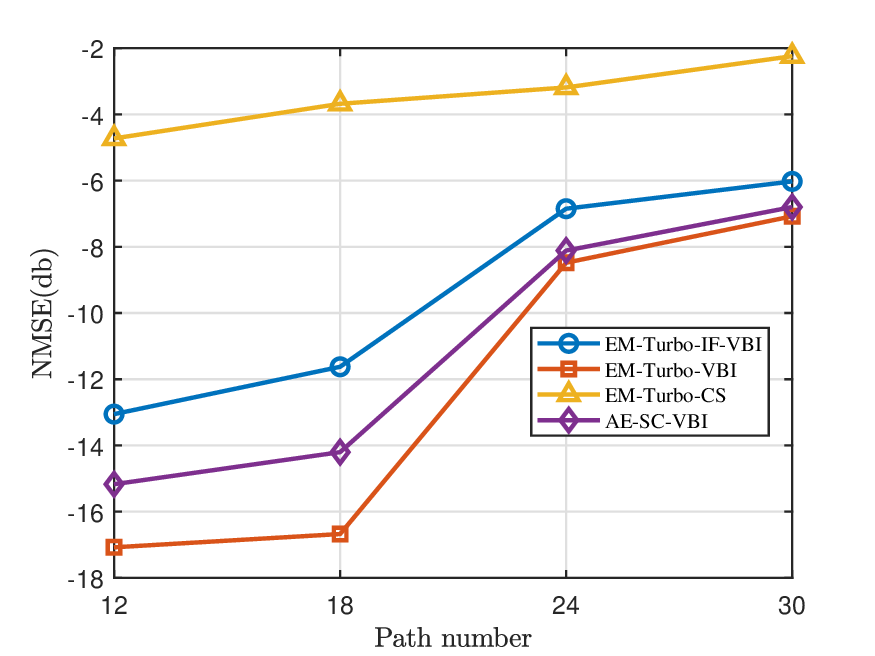}}
\par\end{centering}
\textcolor{blue}{\caption{{\small{}\label{fig:path number}NMSE of channel estimation versus
the} number of paths.}
}
\end{figure}

In Fig. \ref{fig:path number}, we focus on how the number of channel
paths affects the channel estimation performance, where the compression
ratio is set to 8 and the SNR is set to 5 dB. It is clear to see that
the channel estimation performance of all algorithms gets worse as
the number of paths increases. Due to the compression ratio is larger
than 4, the EM-Turbo-CS algorithm works poorly. Again, our proposed
AE-SC-VBI achieve a significant performance gain over the EM-Turbo-IF-VBI.\textcolor{blue}{}

\subsection{Influence of Sparse Prior Model}

\begin{figure}
\begin{centering}
\textcolor{blue}{\includegraphics[width=80mm]{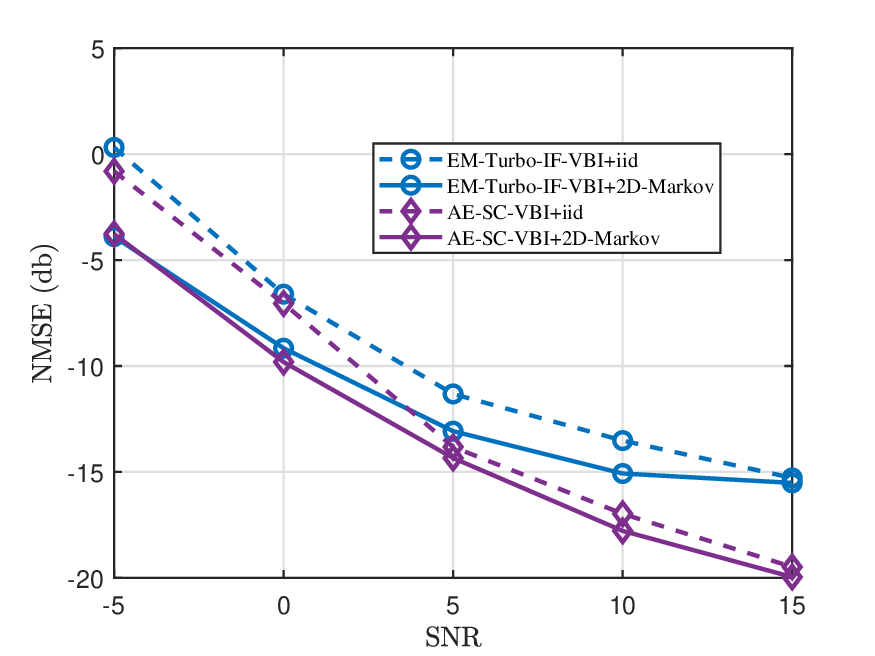}}
\par\end{centering}
\textcolor{blue}{\caption{{\small{}\label{fig:sparse model}NMSE of channel estimation in different
prior model}}
}
\end{figure}

In Fig. \ref{fig:sparse model}, we study the impact of different
sparse prior models, i.e, the proposed 2D-Markov prior model and the
i.i.d. prior model, where the compression ratio is set to 6. In the
low SNR regions, the algorithm with the 2D-Markov prior has a significant
performance gain over the same algorithm with an i.i.d. prior. While
in the high SNR region, the performance gain becomes relatively small.
This is because the observation information is sufficient to estimate
the channel well in high SNR regions. In this case, the structured
prior information contributes less to the channel estimation performance.
\textcolor{blue}{}

\subsection{Influence of Grid Update}

\begin{figure}
\begin{centering}
\textcolor{blue}{\includegraphics[width=80mm]{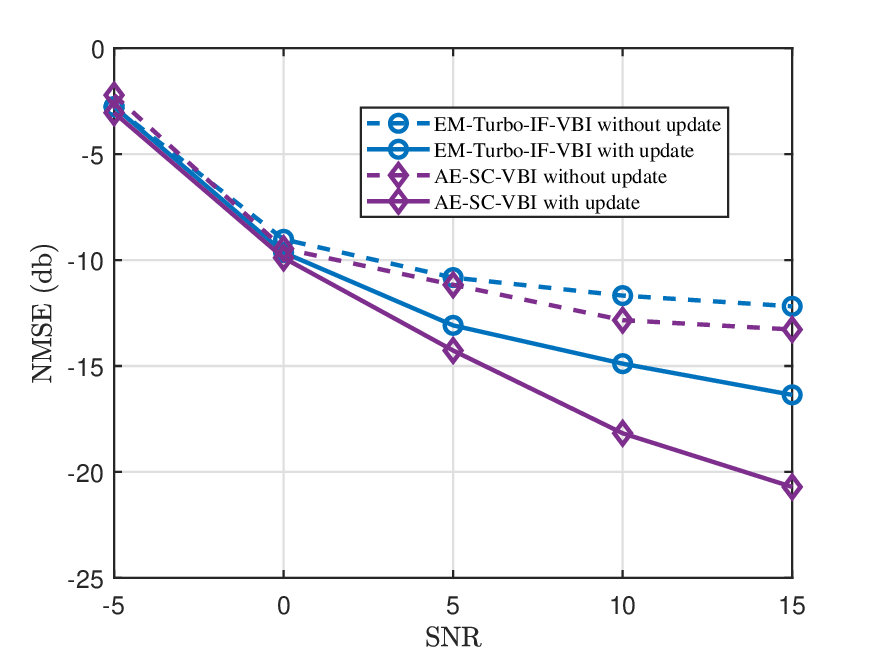}}
\par\end{centering}
\textcolor{blue}{\caption{{\small{}\label{fig:grid_update}NMSE of channel estimation versus
SNR with and without dynamic grid update}}
}
\end{figure}

In Fig. \ref{fig:grid_update}, we study the impact of dynamic grid
update, with the compression ratio set to 6. Our proposed AE-SC-VBI
algorithm updates the grid in the grid estimation module \ref{subsec:The-Grid-Estimation},
while the EM-based algorithms update the grid during the M-step. It
is clear to see that incorporating dynamic grid update leads to a
significant performance improvement in high SNR regions, for both
the proposed AE-SC-VBI algorithm and the EM-Turbo-IF-VBI algorithm.
Note that since the influence of grid update on the EM-Turbo-VBI and
EM-Turbo-CS is similar, their NMSE curves are not shown in Fig. \ref{fig:grid_update}
to avoid too many curves in a single plot.\textcolor{blue}{}

\section{\textcolor{black}{Conclusion\label{sec:Conlusion}}}

\textcolor{black}{We propose an alternating estimation (AE) framework
with a novel SC-VBI algorithm (}AE-SC-VBI\textcolor{black}{) for robust
recovery of structured sparse signals with dynamic grid parameters
in the sensing matrix. The AE framework iterates among three modules
until convergence: the subspace constrained variational Bayesian inference}\textbf{\textcolor{black}{{}
}}\textcolor{black}{(SC-VBI)}\textbf{ }\textcolor{black}{module},
the grid estimation (GE) module and the structured sparse inference
(SSI) module. For fixed sparse signal from the SC-VBI module, the
GE module refines the dynamic gird by directly maximizing the likelihood
function based on the gradient ascent method with backtracking line
search. For fixed extrinsic message from the SC-VBI module, the SSI
module performs sum-product message passing over the structured sparse
prior to exploit the specific sparse structures in practical applications.
For fixed grid from the GE module and extrinsic message from the SSI
module, the SC-VBI module computes a Bayesian estimation of the sparse
signal\textcolor{black}{. In the proposed SC-VBI algorithm, the high-dimensional
matrix inverse is replaced by a low-dimensional subspace constrained
matrix inverse. We further establish the convergence of the SC-VBI
to a stationary solution of the KLD minimization problem. Finally,
we apply the proposed }AE-SC-VBI algorithm\textcolor{black}{{} to solve
a massive MIMO channel estimation problem. The proposed }AE-SC-VBI\textcolor{black}{{}
algorithm is shown in the simulations to achieve significant gains
over baseline algorithms.}

\appendix

\subsection{\textcolor{black}{Proof of Lemma \ref{lem:ICVBI}\label{subsec:Proof-of-lemmaic}}}

Given $q\left(\boldsymbol{\rho}\right),q\left(\boldsymbol{s}\right),q\left(\kappa\right)$,
the KLD can be calculated as
\begin{align}
\textrm{KL}\left(q\parallel p\right)= & \int q\left(\boldsymbol{v}\right)\ln\frac{q\left(\boldsymbol{v}\right)}{\hat{p}\left(\boldsymbol{v}\mid\boldsymbol{y}\right)}\textrm{d}\boldsymbol{v}\nonumber \\
= & \left\langle \ln q\left(\boldsymbol{x}\right)\right\rangle _{q\left(\boldsymbol{x}\right)}-\left\langle \ln\hat{p}\left(\boldsymbol{v},\boldsymbol{y}\right)\right\rangle _{q\left(\boldsymbol{v}\right)}+C\nonumber \\
= & \left\langle \ln q\left(\boldsymbol{x}\right)\right\rangle _{q\left(\boldsymbol{x}\right)}-\left\langle \ln p\left(\boldsymbol{y}\mid\boldsymbol{x},\kappa\right)\right\rangle _{q\left(\boldsymbol{x}\right)q\left(\kappa\right)}\nonumber \\
 & -\left\langle \ln p\left(\boldsymbol{x}\mid\boldsymbol{\rho}\right)\right\rangle _{q\left(\boldsymbol{x}\right)q\left(\boldsymbol{\rho}\right)}+C,\nonumber \\
= & \left\langle \ln q\left(\boldsymbol{x}\right)\right\rangle _{q\left(\boldsymbol{x}\right)}-\left\langle \kappa\right\rangle \left\langle \left\Vert \boldsymbol{y}-\mathbf{A}\boldsymbol{x}\right\Vert ^{2}\right\rangle _{q\left(\boldsymbol{x}\right)}\boldsymbol{\mu}^{H}\nonumber \\
 & -\left\langle \boldsymbol{x}^{H}\textrm{diag}\left(\left\langle \boldsymbol{\rho}\right\rangle \right)\boldsymbol{x}\right\rangle _{q\left(\boldsymbol{x}\right)}+C,\label{eq:KLD_q(x)}
\end{align}
where $C$ is a constant. Based on the constraint in (\ref{eq:GScon}),
Problem$\mathcal{\mathscr{A}}_{\mathrm{VBI}}$ is equivalent to finding
the optimal parameters of $q\left(\boldsymbol{x}\right)$, denoted
by $\left\{ \ddot{\boldsymbol{\mu}},\ddot{\boldsymbol{\sigma}}^{2}\right\} $,
so that the KLD in (\ref{eq:KLD_q(x)}) is minimized. The equivalent
optimization problem is formulated as
\begin{align}
\ddot{\boldsymbol{\mu}},\ddot{\boldsymbol{\sigma}}^{2}= & \underset{\boldsymbol{u},\boldsymbol{\sigma}^{2}}{\text{min}}\left[-\sum_{n=1}^{N}\ln\sigma_{n}^{2}-\boldsymbol{\mu}^{H}\textrm{diag}\left(1/\boldsymbol{\sigma}^{2}\right)\boldsymbol{\mu}\right.\nonumber \\
 & -\left\langle \left(\boldsymbol{x}-\boldsymbol{\mu}\right)^{H}\textrm{diag}\left(1/\boldsymbol{\sigma}^{2}\right)\left(\boldsymbol{x}-\boldsymbol{\mu}\right)\right\rangle _{q\left(\boldsymbol{x}\right)}\nonumber \\
 & \left.+\left\langle \kappa\right\rangle \left\langle \left\Vert \boldsymbol{y}-\mathbf{A}\boldsymbol{x}\right\Vert ^{2}\right\rangle _{q\left(\boldsymbol{x}\right)}+\left\langle \boldsymbol{x}^{H}\textrm{diag}\left(\left\langle \boldsymbol{\rho}\right\rangle \right)\boldsymbol{x}\right\rangle _{q\left(\boldsymbol{x}\right)}\right]\nonumber \\
= & \underset{\boldsymbol{u},\boldsymbol{\sigma}^{2}}{\text{min}}\left[-\sum_{n=1}^{N}\ln\sigma_{n}^{2}-\boldsymbol{\mu}^{H}\textrm{diag}\left(1/\boldsymbol{\sigma}^{2}\right)\boldsymbol{\mu}\right.\nonumber \\
 & +\left\langle \boldsymbol{x}^{H}\left(\mathbf{W}-\textrm{diag}\left(1/\boldsymbol{\sigma}^{2}\right)\right)\boldsymbol{x}\right\rangle _{q\left(\boldsymbol{x}\right)}\nonumber \\
 & \left.-\left\langle 2\mathfrak{Re}\left\{ \boldsymbol{x}^{H}\left(\boldsymbol{b}-\textrm{diag}\left(1/\boldsymbol{\sigma}^{2}\right)\boldsymbol{\mu}\right)\right\} \right\rangle _{q\left(\boldsymbol{x}\right)}\right]\nonumber \\
= & \underset{\boldsymbol{u},\boldsymbol{\sigma}^{2}}{\text{min}}\left[-\sum_{n=1}^{N}\ln\sigma_{n}^{2}-\boldsymbol{\mu}^{H}\textrm{diag}\left(1/\boldsymbol{\sigma}^{2}\right)\boldsymbol{\mu}\right.\nonumber \\
 & +\boldsymbol{\mu}^{H}\left(\mathbf{W}-\textrm{diag}\left(1/\boldsymbol{\sigma}^{2}\right)\right)\boldsymbol{\mu}\nonumber \\
 & +\textrm{Tr}\left(\left(\mathbf{W}-\textrm{diag}\left(1/\boldsymbol{\sigma}^{2}\right)\right)\textrm{diag}\left(\boldsymbol{\sigma}^{2}\right)\right)\nonumber \\
 & \left.-2\mathfrak{Re}\left\{ \boldsymbol{\mu}^{H}\left(\boldsymbol{b}-\textrm{diag}\left(1/\boldsymbol{\sigma}^{2}\right)\boldsymbol{\mu}\right)\right\} \right]\nonumber \\
= & \underset{\boldsymbol{u},\boldsymbol{\sigma}^{2}}{\text{min}}\left[\boldsymbol{u}^{H}\mathbf{W}\boldsymbol{u}-2\mathfrak{Re}\left\{ \boldsymbol{u}^{H}\boldsymbol{b}\right\} +\sum_{n=1}^{N}\left(W_{n}\sigma_{n}^{2}-\ln\sigma_{n}^{2}\right)\right],
\end{align}
which is equal to (\ref{eq:EQIC}).

\subsection{Proof of Theorem \ref{thm:Convergence-of-SC-VBI} \label{subsec:Proof-of-TheoremSCVBI}}

The SC-VBI can be viewed as an alternating optimization method to
solve Problem $\mathcal{\mathscr{A}}_{\mathrm{VBI}}$. It is clear
that the SC-VBI can monotonically decreasing the KLD objective, thus
the KLD will converge to a limit. For convenience, we use $\boldsymbol{\xi}\triangleq\left\{ \boldsymbol{\mu},\boldsymbol{\sigma}^{2},\widetilde{\boldsymbol{a}},\widetilde{\boldsymbol{b}},\widetilde{\boldsymbol{\lambda}},\widetilde{c},\widetilde{d}\right\} $
to represent the parameters of $q\left(\boldsymbol{v}\right)$. Let
$\boldsymbol{\xi}_{j}$ denote the $j\textrm{-th}$ block of $\boldsymbol{\xi}$
for $j=1,\ldots,B$, where $B=\left|\boldsymbol{\xi}\right|$ . Then,
the KLD in (\ref{eq:KLDmin}) can be rewritten as a function of $\boldsymbol{\xi}$,
i.e.
\begin{equation}
\textrm{KL}\left(\boldsymbol{\xi}\right)=\int q\left(\boldsymbol{v};\boldsymbol{\xi}\right)\ln\frac{q\left(\boldsymbol{v};\boldsymbol{\xi}\right)}{\hat{p}\left(\boldsymbol{v}\mid\boldsymbol{y};\boldsymbol{\xi}\right)}\textrm{d}\boldsymbol{v}.
\end{equation}
In the following, we will prove that every limiting point $q^{*}\left(\boldsymbol{v};\boldsymbol{\xi}^{*}\right)$
generated by the SC-VBI is a stationary solution of Problem $\mathcal{\mathscr{A}}_{\mathrm{VBI}}$.

In the $i\textrm{-th}$ iteration of the SC-VBI, we update $\boldsymbol{\xi}_{j},\forall j$
alternatively as
\begin{align}
\boldsymbol{\xi}_{j}^{\left(i\right)} & \begin{cases}
=\underset{\boldsymbol{\xi}_{j}}{\textrm{argmin}}\textrm{KL}\left(\boldsymbol{\xi}_{j},\boldsymbol{\xi}_{-j}^{\left(i-1\right)}\right), & \textrm{if}\ \boldsymbol{\xi}_{j}\neq\boldsymbol{\mu},\\
=\boldsymbol{\mu}^{\left(i-1\right)}-\epsilon_{x}^{\left(i\right)}\nabla_{\boldsymbol{\mu}}\textrm{KL}\left(\boldsymbol{\mu},\boldsymbol{\xi}_{-j}^{\left(i-1\right)}\right)\mid_{\boldsymbol{\mu}=\boldsymbol{\mu}^{\left(i-1\right)}}, & \textrm{if}\ \boldsymbol{\xi}_{j}=\boldsymbol{\mu},
\end{cases}\label{eq:update_rule}
\end{align}
where $\left(\cdotp\right)^{\left(i\right)}$ stands for the $i\textrm{-th}$
iteration and $\boldsymbol{\xi}_{-k}^{\left(i\right)}=\left(\boldsymbol{\xi}_{1}^{\left(i\right)},\ldots,\boldsymbol{\xi}_{k-1}^{\left(i\right)},\boldsymbol{\xi}_{k+1}^{\left(i-1\right)},\ldots,\boldsymbol{\xi}^{\left(i-1\right)}\right)$.
For case of $\boldsymbol{\xi}_{j}\neq\boldsymbol{\mu}$, it is clear
that $\textrm{KL}\left(\boldsymbol{\xi}_{j},\boldsymbol{\xi}_{-j}^{\left(i-1\right)}\right)$
is minimized w.r.t. $\boldsymbol{\xi}_{j}$. While for case of $\boldsymbol{\xi}_{j}=\boldsymbol{\mu}$,
we have $\textrm{KL}\left(\boldsymbol{\mu}^{\left(i-1\right)},\boldsymbol{\xi}_{-j}^{\left(i-1\right)}\right)\geq\textrm{KL}\left(\boldsymbol{\mu}^{\left(i\right)},\boldsymbol{\xi}_{-j}^{\left(i-1\right)}\right)$
based on the property of gradient update, where the equality holds
only when the gradient w.r.t. $\boldsymbol{\mu}$ is zero. Therefore,
the KLD will keep decreasing until converging to a certain value,
and we must have 
\begin{equation}
\lim_{i\rightarrow\infty}\nabla_{\boldsymbol{\xi}_{j}}\textrm{KL}\left(\boldsymbol{\xi}_{j},\boldsymbol{\xi}_{-j}^{\left(i-1\right)}\right)=0,\forall j.\label{eq:gradzero}
\end{equation}
 (otherwise, the KLD will keep decreasing to negative infinity, which
contradicts with the fact that $\textrm{KL}\left(\boldsymbol{\xi}\right)\geq0$).
Then according to (\ref{eq:gradzero}) and the property of gradient
update, we must have $\lim_{i\rightarrow\infty}\left\Vert \boldsymbol{\mu}^{\left(i\right)}-\boldsymbol{\mu}^{\left(i-1\right)}\right\Vert =0$.
Moreover, it follows from (\ref{eq:gradzero}) and the strong convexity
of $\textrm{KL}\left(\boldsymbol{\xi}_{j},\boldsymbol{\xi}_{-j}^{\left(i-1\right)}\right)$
w.r.t. $\boldsymbol{\xi}_{j},\forall\boldsymbol{\xi}_{j}\neq\boldsymbol{\mu}$
that $\lim_{i\rightarrow\infty}\left\Vert \boldsymbol{\xi}_{j}^{\left(i\right)}-\boldsymbol{\xi}_{j}^{\left(i-1\right)}\right\Vert =0,\forall\boldsymbol{\xi}_{j}\neq\boldsymbol{\mu}$.
Therefore, we have
\begin{equation}
\lim_{i\rightarrow\infty}\left\Vert \boldsymbol{\xi}_{j}^{\left(i\right)}-\boldsymbol{\xi}_{j}^{\left(i-1\right)}\right\Vert =0,\forall j.\label{eq:continuexi}
\end{equation}
It follows from (\ref{eq:continuexi}) that all the $B$ sequences
$\left\{ \boldsymbol{\xi}_{j}^{(i)},\boldsymbol{\xi}_{-j}^{(i)}\right\} ,j=0,1,...,B-1$
have the same set of limiting points. Let $\left\{ \boldsymbol{\xi}_{j}^{(i_{t})},\boldsymbol{\xi}_{-j}^{(i_{t})},t=1,2,...\right\} $
denote a subsequence that converges to a limiting point $\boldsymbol{\xi}^{*}$.
Suppose $\boldsymbol{\xi}^{*}$ is not a stationary point of $\textrm{KL}\left(\boldsymbol{\xi}\right)$,
then $\nabla_{\boldsymbol{\xi}}\textrm{KL}\left(\boldsymbol{\xi}\right)\neq0$
and it follows from (\ref{eq:continuexi}) that $\lim_{t\rightarrow\infty}\nabla_{\boldsymbol{\xi}_{j}}\textrm{KL}\left(\boldsymbol{\xi}_{j},\boldsymbol{\xi}_{-j}^{\left(i_{t}\right)}\right)\neq0$
must hold at least for some $j$, which contradicts with (\ref{eq:gradzero}).
Therefore, every limiting point $\boldsymbol{\xi}^{*}$ must be a
stationary point of $\textrm{KL}\left(\boldsymbol{\xi}\right)$. In
other word, every limiting point $q^{*}\left(\boldsymbol{v};\boldsymbol{\xi}^{*}\right)$
generated by the SC-VBI is a stationary solution of Problem $\mathcal{\mathscr{A}}_{\mathrm{VBI}}$.


\end{document}